\documentclass[]{interact}

\usepackage{epstopdf}
\usepackage{subfigure}

\usepackage[numbers,sort&compress]{natbib}
\bibpunct[, ]{[}{]}{,}{n}{,}{,}
\renewcommand\bibfont{\fontsize{10}{12}\selectfont}
\makeatletter
\def\NAT@def@citea{\def\@citea{\NAT@separator}}
\makeatother

\theoremstyle{plain}

\theoremstyle{definition}

\theoremstyle{remark}

\begin{document}

\articletype{QMUL-PH-17-15}

\title{The double copy: gravity from gluons}

\author{ \name{C.~D. White\textsuperscript{a}\thanks{Email:
      christopher.white@qmul.ac.uk}} \affil{\textsuperscript{a}Centre
    for Research in String Theory, School of Physics and Astronomy,
    Queen Mary University of London, 327 Mile End Road, London E1 4NS,
    UK} }

\maketitle

\begin{abstract}
Three of the four fundamental forces in nature are described by
so-called {\it gauge theories}, which include the effects of both
relativity and quantum mechanics. Gravity, on the other hand, is
described by General Relativity, and the lack of a well-behaved
quantum theory - believed to be relevant at the centre of black holes,
and at the Big Bang itself - remains a notorious unsolved
problem. Recently a new correspondence, the {\it double copy}, has
been discovered between scattering amplitudes (quantities related to
the probability for particles to interact) in gravity, and their gauge
theory counterparts. This has subsequently been extended to other
quantities, providing gauge theory analogues of e.g. black holes. We
here review current research on the double copy, and describe some
possible applications.
\end{abstract}


\section{Introduction}
\label{sec:intro}

Fundamental physics continues to seek answers to the biggest questions
facing humankind: where did the universe come from? What is it made
of?  How will it end? The past few hundreds of years of development
have combined precision experiments with abstract theoretical
reasoning, and our current understanding of the basic building blocks
of nature is as follows. The universe contains {\it matter}, which is
acted upon by {\it forces}. All observed forces are consequences of
only four {\it fundamental forces}. Three of these - electromagnetism,
the weak nuclear force, and the strong nuclear force - are described
by the Standard Model (SM) of particle physics, which also lists all
known matter particles -- quarks and leptons -- that combine to make
the various composite particles observed in nature. All forces and
matter types in the SM are described by fields filling all of
spacetime. The equations describing these fields include the effects
of quantum mechanics and relativity, so that the SM is an example of a
{\it quantum field theory} (QFT). Indeed, the particles themselves
emerge as quanta of the fields, with the canonical example being the
photon (a quantum of the electromagnetic field). Furthermore, the
fields have certain abstract mathematical symmetries, so that this
type of QFT is also called a {\it (non)-Abelian gauge theory}. We will
define these terms more precisely in what follows.

\begin{figure}
\begin{center}
\scalebox{0.6}{\includegraphics{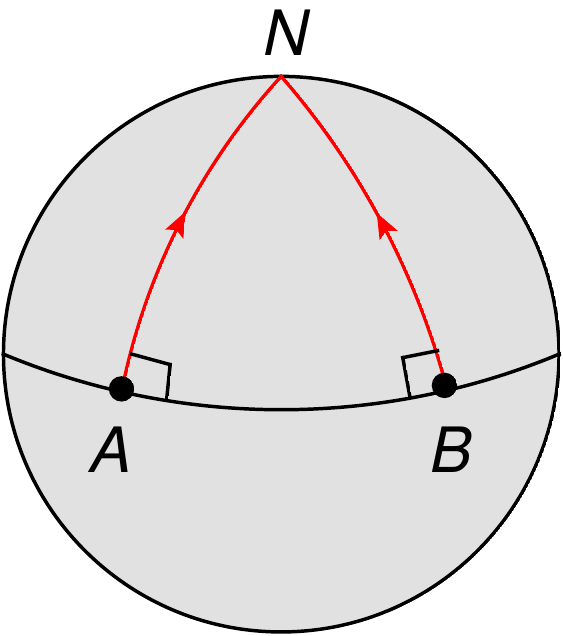}}
\caption{On a (spherical) curved space, observers at $A$ and $B$ are
  told to walk towards the north pole ($N$), and start off mutually
  parallel. Due to the curvature, the observers will move towards each
  other, which looks like an attractive force.}
\label{fig:sphere}
\end{center}
\end{figure}

The fourth force in nature is gravity, and is currently best described
by the General Theory of Relativity (GR). This tells us that space and
time are dynamical, rather than being a fixed background upon which
particles and forces operate. The equations of GR tell us that matter
and energy warp the fabric of spacetime in a prescribed way, such that
this curvature can be identified with the the gravitational force. A
simple analogy for how curvature can produce an attractive force is
shown in figure~\ref{fig:sphere}. 
GR underpins much of our understanding of how the universe works at
very large scales. It tells us that black holes may exist, for
example, and gives rise to the possibility that the universe may have
expanded outwards from a ``big bang'' at some finite time in the
past. It also predicts the existence of {\it gravitational waves}:
ripples in the fabric of spacetime that are analogous to the wave-like
solutions of Maxwell's equations for electromagnetism, and which have
only recently been observed for the first
time~\cite{Abbott:2016blz}. On a more practical level, one must
account for the curvature of spacetime in satellite communications,
and GR has even become part of everyday life through the use of
positioning systems in smartphones.

Despite the wide-ranging success of both the SM and GR, many puzzles
remain. It is not known, for example, why there are only four
fundamental forces, and why the various matter particles have their
particular properties (e.g. masses, and charges with respect to each
force). It is not fully understood why matter dominated over
antimatter in the early universe, and present day astrophysical
observations imply the existence of both {\it dark matter} and {\it
  dark energy}, neither of which is present in the SM. Furthermore,
the classical theory of GR breaks down at extreme points in spacetime,
such as at the centre of a black hole, or the big bang itself. In such
places, the curvature of spacetime becomes infinite, which is not
physically sensible. It is thus widely thought that both the SM and GR
are part of some larger theoretical framework, which may include
quantum effects for the gravitational force, in line with the other
forces. One may attempt to turn GR into a quantum field theory, such
that gravity is carried by a {\it graviton}, analagous to the photon
that carries the electromagnetic force in the SM. Such attempts fail,
however, due to the fact that calculations become infinite when
gravitons are emitted and absorbed at arbitrarily short distances,
corresponding to high energy scales. These so-called {\it ultraviolet
  (UV) divergences} are also present in the SM, but can be removed by
redefinitions of the fields and parameters entering the theory, such
as particle masses, and the {\it coupling constants} describing the
intrinsic strengths of each force. This is known as {\it
  renormalisation}, and the same procedure does not work for GR, such
that the theory is described as {\it non-renormalisable}.

The apparent inability to reconcile quantum mechanics with the theory
of gravity is one of the most notorious open problems in theoretical
physics. Quantum gravity may resolve the inability of GR to fully
describe black hole physics, the big bang, or dark energy. There are a
number of possibilities for such a theory. Firstly, it may turn out
that GR is alright after all if thought about in the right way
e.g. calculations that rely purely on {\it perturbation theory} (an
expansion in the strength of the gravitational force) may be
insufficient. One such proposal is the asymptotic safety idea of
ref.~\cite{Weinberg:1980gg}. Secondly, there may be a modified field
theory of gravity, that is renormalisable in perturbation theory. An
open possibility is ${\cal N}=8$ Supergravity, which has additional
matter content alongside the graviton, in a highly constrained way
such that a certain symmetry between bosons and fermions ({\it
  supersymmetry}) is made manifest. Thirdly, quantum gravity may not
be a field theory that gives rise to particles, but a different type
of theory. One such proposal is {\it string theory}, in which both
gauge theories and gravity emerge from the dynamics of vibrating
strings. Of course, a fourth possibility is that there is no quantum
theory of gravity at all, but this would leave significant unsolved
puzzles in both the SM and GR.

One of the main problems in investigating quantum gravity is the sheer
complexity of the calculations involved, which quickly become
intractable even with the aid of powerful computers. It seems that new
methods and insights are needed, and one such technique has arisen in
the last few years: {\it the double
  copy}~\cite{Bern:2010ue,Bern:2010yg}. This relates quantities
calculated in a gauge theory, with similar quantities obtained in a
gravity theory. The original form of this correspondence involved
scattering amplitudes - complex-valued functions of momenta which are
related to the probability for a given set of particles to
interact. However, it has since been extended to other types of object
in gauge and gravity theories - such as exact classical solutions,
including black holes. Furthermore, similar correspondences have been
found between other types of field theory, with or without
supersymmetry.

The double copy has the potential to revolutionise our understanding
of gravity, given that it relates theories like those in the SM, whose
quantum behaviour we understand well, with gravity. It suggests that,
if one thinks about gravity in the right way, it is much simpler than
traditional calculations in GR would seem to suggest. We also get new
insights into gauge theory itself. For the double copy between
scattering amplitudes in gauge and gravity theories to work, the gauge
theory results have to be written to obey an intriguing symmetry
between the parts relating to the charges of each gluon, and the parts
relating to their momenta, polarisations etc. This is known as {\it
  BCJ duality}~\cite{Bern:2008qj}, and implies that these various
degrees of freedom are much more closely related than previously
thought. In order to be able to define the double copy more formally,
let us first study gauge theories in more detail. 

\section{Gauge theories}
\label{sec:gauge}

Here, we review the basic ideas underlying gauge theories - the type
of quantum field theory that describes three out of the four
fundamental forces in nature, as contained in the Standard Model of
particle physics. Arguably the most familiar gauge theory is that of
electrodynamics, which we turn to first.

\subsection{Quantum electrodynamics (QED)}
\label{sec:EM}

The basic idea of quantum field theory is that matter and forces are
described by {\it fields} filling all of spacetime. There are
equations of motion for these fields, which can have wavelike
solutions. In the quantum theory, waves of a given frequency $\nu$
cannot have arbitrary energy, but instead come as discrete {\it
  quanta}, with energy $E=h\nu$, where $h$ is Planck's constant. The
canonical example of this is the electromagnetic field: this is
described by Maxwell's equations, whose classical wavelike solutions
constitute the electromagnetic spectrum. The quanta of this field are
called {\it photons}, which are often referred to as the particles
that carry the electromagnetic force. It is important to bear in mind,
however, that the particles of quantum field theory are not quite the
same as particles in Newtonian physics: QFT contains both wavelike and
particle behaviour, such that which aspect appears in any given
experiment depends on what is being measured. This is the famous
property of {\it wave-particle duality}.

\begin{figure}
\begin{center}
\scalebox{0.8}{\includegraphics{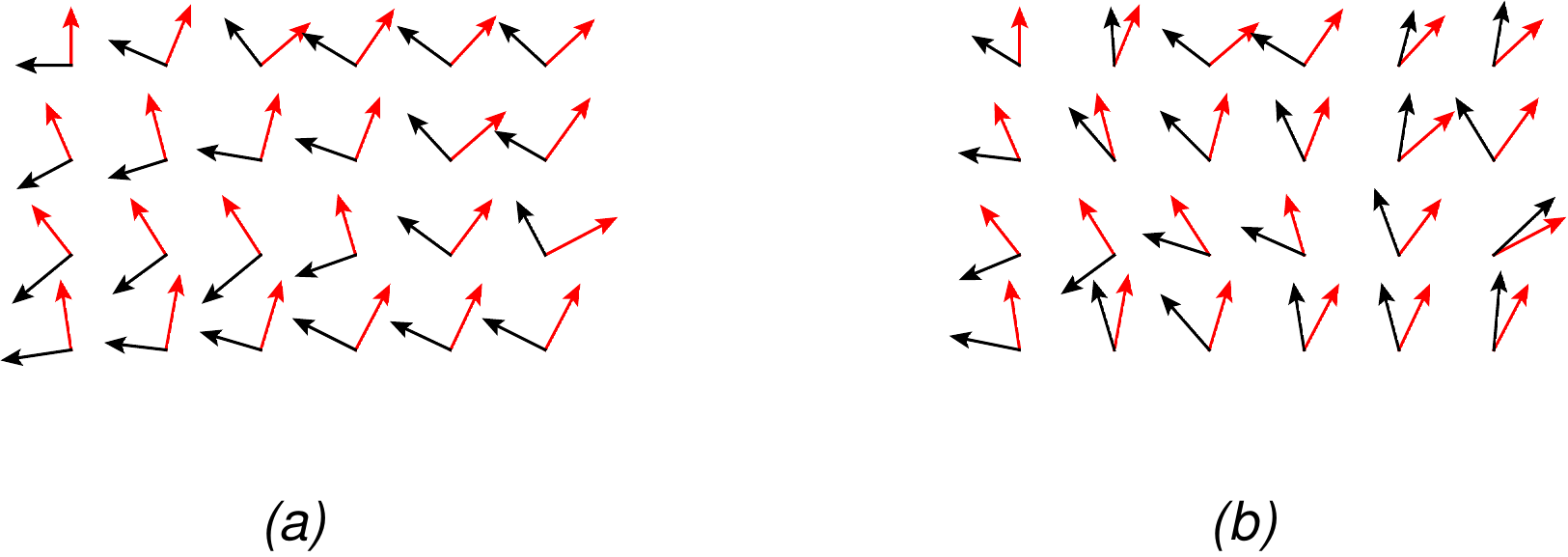}}
\caption{Schematic representation of (a) a global gauge
  transformation, corresponding to simultaneously shifting the phase
  of a field at all spacetime points by the same amount (here $\pi/2$,
  or $90^\circ$); (b) a local gauge transformation, in which the phase
  is shifted by different amounts at each point.}
\label{fig:gaugetrans}
\end{center}
\end{figure}

Similarly, matter is also described by fields. The electron, for
example, is described by a spinor field $\Psi(x^\mu)$, where
$x^\mu=(t,\vec{x})$ denotes an arbitrary point in
spacetime~\footnote{Throughout this review, we work in {\it natural
    units} such that $\hbar=c=1$, where $\hbar=h/2\pi$, and $c$ is the
  speed of light.}. If you have not seen spinors before, this is
simply a mathematical object with four components, such that it has
enough degrees of freedom to represent the two possible spin
components of the electron (which is a fermion), or its antiparticle
(the positron). Furthermore, the field is complex-valued. The equation
of motion for this field is the famous {\it Dirac
  equation}~\cite{Dirac:1928hu}, and it involves combinations of the
field $\Psi$ and its complex conjugate, such that any observable
properties of electrons are real numbers, as they should be. In turn,
this means that the equations of the electron are invariant under the
transformation
\begin{equation}
\Psi(x^\mu)\rightarrow e^{i\alpha}\Psi(x^\mu)
\label{gauge1}
\end{equation}
for some constant parameter $\alpha$, as the factor $e^{i\alpha}$ will
be cancelled out by a factor $e^{-i\alpha}$ in the complex conjugate
field. We can visualise this transformation as follows. If the
electron field is complex-valued, it has a magnitude and a phase at
all points in spacetime. At each point, we can represent the phase by
an arrow of unit magnitude, pointing in some direction on the unit
circle. The transformation of eq.~(\ref{gauge1}) then corresponds to
rotating all arrows by the same amount simultaneusly at all spacetime
points, as shown in figure~\ref{fig:gaugetrans}(a), in which the red
arrows are obtained by rotating all the black arrows by $\pi/2$, or
$90^\circ$.
This is called a {\it global gauge transformation} and it is clear
that the theory must be invariant under such operations: we can always
choose to redefine what we mean by the zero of phase, so that only
{\it differences} in phase are physically meaningful. A global gauge
transformation amounts to such a redefinition of the zero of phase.

In fact, QED turns out to have a much more remarkable symmetry than
this. The equations of the theory are invariant under a generalisation
of eq.~(\ref{gauge1}), in which the phase factor $\alpha$ itself
depends on spacetime position:
\begin{equation}
\Psi(x^\mu)\rightarrow e^{i\alpha(x^\mu)}\Psi(x^\mu).
\label{gauge2}
\end{equation}
In the geometric picture discussed above, eq.~(\ref{gauge2})
corresponds to rotating the arrows describing the phase of the field
by different amounts at different spacetime points, as depicted in
figure~\ref{fig:gaugetrans}(b). This is known as a {\it local gauge
  transformation}, and it is not at all obvious a priori that this
should be a symmetry of nature. Indeed, local gauge transformations
are infinitely richer than global ones, given that there is a separate
phase factor for every point in spacetime. Curiously, in order to make
the equations describing the electron field invariant under local
gauge transformations, one must introduce a 4-vector valued field
$A^\mu(x^\mu)$, that couples to the electron in a prescribed way. This
field turns out to correspond exactly to the known 4-vector potential
in electromagnetism, whose components are
\begin{equation}
A^\mu=(\phi,\vec{A}),
\label{Amudef}
\end{equation}
with $\phi$ and $\vec{A}$ the electrostatic and magnetic vector
potential respectively. This is a remarkable result: requiring an
abstract mathematical symmetry of the electron equations (local gauge
invariance) necessarily predicts the existence of electromagnetism,
whose equations were first arrived at after hundreds of years of
detailed experiments, involving a plethora of different phenomena!

The above discussion immediately begs the question of whether similar
abstract symmetries can be used to explain the other forces in the SM,
and indeed the answer to this question is yes. Before seeing how this
works, let us look at the transformation of eq.~(\ref{gauge2}) more
formally. Symmetries are described by the branch of mathematics known
as {\it group theory}. Because the transformations of
eq.~(\ref{gauge2}) involve a continuous parameter ($\alpha$), they
form a continuous group, also known as a {\it Lie group}. The
particular group in question is usually denoted $U(1)$, which stands
for the set of unitary $1\times 1$ matrices i.e. numbers $U$
satisfying
\begin{equation}
U^\dagger U=U U^\dagger=1.
\label{unitary}
\end{equation}
This already hints at how one can generalise the above gauge symmetry
to describe other forces in nature: one can look for invariance under
abstract phase transformations corresponding to more complicated Lie
groups. The group $U(1)$ is especially simple, given that the
transformations commute with each other e.g.
\begin{equation}
e^{i\alpha_1}\,e^{i\alpha_2}=e^{i\alpha_2}\,e^{i\alpha_1}.
\label{commute}
\end{equation}
Groups where the elements commute are called {\it abelian}. The groups
describing the other forces in the SM do not have this property: they
are {\it non-abelian}.

\subsection{Non-abelian gauge theories}
\label{sec:nonabel}

We have seen that abelian gauge invariance can give rise to theories
like electromagnetism, in which matter fields interact with a photon
field. A famous theorem due to Emmy Noether~\cite{Noether:1918zz}
implies that whenever a physical theory has a symmetry, there must be
a conserved quantity. The property of electrons that is conserved, due
to local gauge invariance, turns out to be the familiar electric
charge $e$. This charge comes in two types, which we conventionally
call {\it positive} and {\it negative}. The other forces in the SM
also have gauge symmetries associated with them, such that the various
matter particles have charges corresponding to each force. These
should not be confused with electric charge, and indeed they can have
different properties. In particular, it is not true that the different
types of charge have only two types.

\begin{figure}
\begin{center}
\scalebox{0.7}{\includegraphics{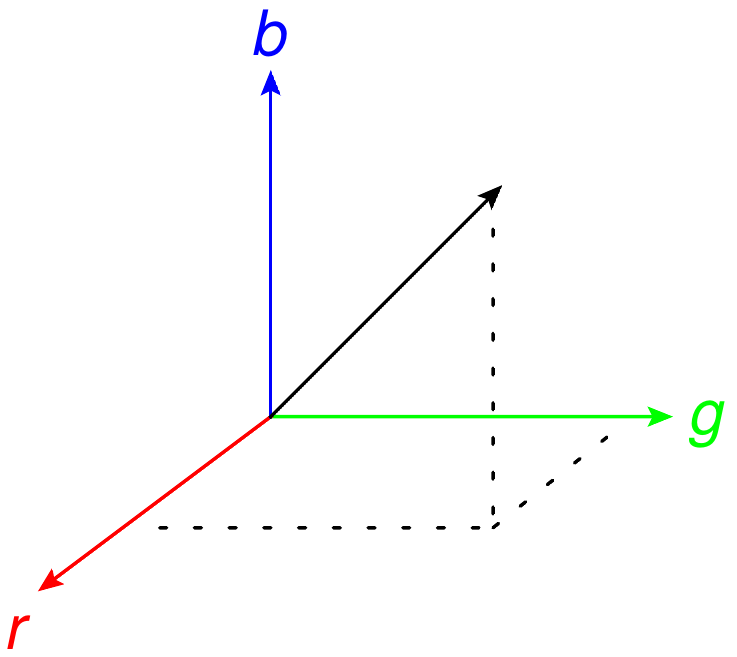}}
\caption{The quark field can be thought of as carrying different
  components, one for each colour. This leads to an abstract vector
  space at each point in spacetime, where the arrow specifies how much
  of each colour is present.}
\label{fig:quarkarrow}
\end{center}
\end{figure}

As an example, let us consider the strong force, felt by quarks
(constituents of the proton and related particles). These feel a
charge under the strong force, which to avoid confusion with the
electric charge is given the name {\it colour}. Unlike electric
charge, colour charge can come in three different types, which are
conventionally labelled {\it red}, {\it green} and {\it blue}, or
$(r,g,b)$ for short. It should be stressed that these are merely
labels - they have nothing to do with visual colours - and that any
other names would have sufficed. Like electrons in QED, quarks are
described by a field. There will be a separate field for each colour
of quark, which we may collect into a vector
\begin{equation}
\psi_{i}(x^\mu)=\left(\psi_r(x^\mu),\psi_g(x^\mu),\psi_b(x^\mu)\right)
\label{quarkdef}
\end{equation}
where $i\in\{r,g,b\}$ is an index labelling the colour. If we want to,
we can visualise this as a single field, but with an abstract vector
at each point in spacetime, that tells us how much redness, greenness
and blueness the field has - see figure~\ref{fig:quarkarrow}.
This reminds us of the arrows in figure~\ref{fig:gaugetrans}, which
described the phase of the electron field at each point. Indeed, it
turns out that the equations of the quark field are invariant under a
local gauge transformation corresponding to rotating the arrow of
figure~\ref{fig:quarkarrow} by arbitrary amounts separately at every
point in spacetime. Physically, this corresponds to a local
redefinition of what we mean by redness, greenness and blueness. As
for QED, requiring this invariance requires introducing an additional
vector field. This additional field describes the {\it gluon}, namely
the particle that carries the strong force.

Let us surmise the Lie group that corresponds to these gauge
transformations. Firstly, moving the arrow in
figure~\ref{fig:quarkarrow} translates to some transformation
\begin{equation}
\psi_i\rightarrow\sum_j U_{ij}\psi_j.
\label{quarktrans}
\end{equation}
As for QED, the equations for the quark field depend only on
combinations of $\psi_i$ with its complex conjugate, so that all
measured quantities are real. Requiring such combinations to be
invariant means that the matrix $U_{ij}$ be {\it
  unitary}. Furthermore, colour charge is conserved, so cannot be
created or destroyed. This means that the length of the arrow in
figure~\ref{fig:quarkarrow} cannot change under
eq.~(\ref{quarktrans}), so $U_{ij}$ must have unit determinant. Thus,
$U_{ij}$ is a so-called {\it special unitary} matrix of dimension
three, and the group of such transformations is called $SU(3)$. Such a
matrix has 8 degrees of freedom, thus demanding local gauge invariance
under all possible $SU(3)$ transformations implies that the gluon
field must have 8 components. This corresponds to the fact that the
gluons themselves carry colour. If, for example, a red quark changes
into a green one by emitting a gluon, the latter must supply an
outgoing red charge and an incoming green one so that the total colour
charge is conserved. There are eight independent combinations of
colour charge that a gluon can carry, so that we usually write the
gluon field as $A^{\mu\,a}(x^\mu)$, where $\mu$ is the spacetime
index, and $a\in\{1,\ldots,8\}$ is the colour index.

A similar mechanism to that explained here can be used to describe all
forces in the SM. For the purposes of this review, we can consider a
more abstract theory, in which we get rid of the quarks alogether, and
consider the gluon field by itself. We may also consider other gauge
groups, other than $SU(3)$. Such a theory is more commonly called {\it
  Yang-Mills theory}, and it is this that enters the simplest form of
the double copy.

\section{Scattering amplitudes}
\label{sec:amps}

\begin{figure}
\begin{center}
\scalebox{0.9}{\includegraphics{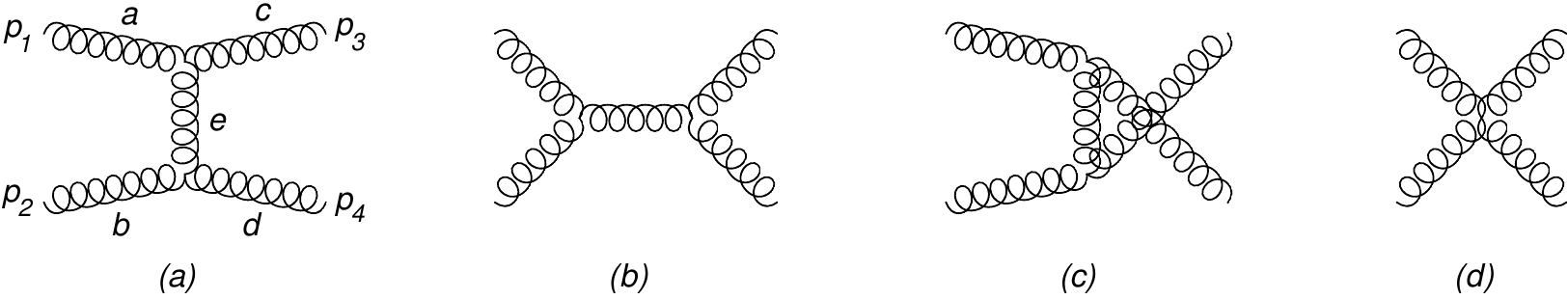}}
\caption{Feynman diagrams for two gluons (shown as curly lines)
  scattering to two gluons: (a) the first gluon emits a gluon that is
  absorbed by the second, or vice versa; (b) two gluons annihilate to
  make a single gluon, which splits again into two; (c) similar to
  process (a), but with the final state momenta interchanged; (d) two
  gluons annihilate and immediately produce two gluons with different
  momenta. In diagram (a), we have labelled the colour index of each
  gluon, and the 4-momenta of the initial and final state gluons.}
\label{fig:Feynman}
\end{center}
\end{figure}

A particularly relevant question to ask in a given quantum field
theory is what happens when particles interact each other. This is
especially topical given that our main way of testing potential new
physics theories is to use {\it particle accelerators}, which collide
beams of particles together, before measuring the resulting debris. A
convenient way to depict particle interactions in a given theory is to
use {\it Feynman diagrams}. Roughly speaking, we can think of these as
space-time diagrams showing the history of a scattering event, where
different types of line represent various types of
particle~\footnote{Feynman diagrams actually represent a
  Lorentz-invariant sum of possible scattering histories, to be
  consistent with special relativity, so should not be thought of too
  literally!}. Some simple examples are shown in
figure~\ref{fig:Feynman}, which shows the different possibilities for
two incoming gluons to produce two outgoing gluons.
Each curly line represents a gluon, which carries an index $a$
corresponding to the colour index in the gauge field
$A_\mu^a$. Furthermore, each incoming or outgoing gluon will have a
4-momentum 
\begin{equation}
p_i^\mu=(E_i,\vec{p}_i),
\label{pidef}
\end{equation}
where $E_i$ and $\vec{p}_i$ are the (relativistic) energy and
3-momentum respectively. The rules of quantum field theory tell us
that, for a given number of incoming and outgoing particles, we must
draw all distinct connected diagrams containing a given set of
vertices, where the latter are determined by the theory. In Yang-Mills
(pure gluon) theory, there are vertices connecting either three or
four gluons at a single point, leading to the four diagrams of
figure~\ref{fig:Feynman}. Note that for graphs (a)--(c), there is an
internal line in addition to the four external gluon lines. This
represents a so-called {\it virtual particle}, that is not observed
directly, but exchanged between incoming or outgoing
particles. Four-momentum is conserved at all vertices. For example, in
diagram (a) we can associate the 4-momentum
\begin{equation}
q^\mu=(p_1-p_3)^\mu
\label{qdef}
\end{equation}
with the internal line, where the minus sign on the right-hand side
arises from the fact that $p_1$ is flowing into the vertex, but $p_3$
is flowing out (we have also chosen $q^\mu$ to flow outwards). The
conservation of 4-momentum at all vertices of a Feynman diagram is
reminiscent of Kirchoff's current law for electrical currents, and can
be applied just as straightforwardly. 

\begin{figure}
\begin{center}
\scalebox{0.6}{\includegraphics{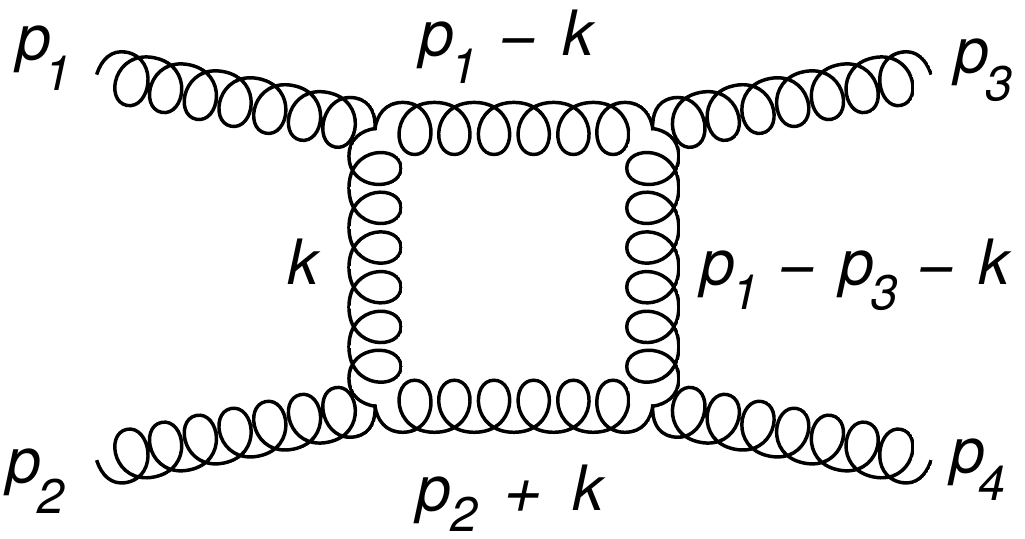}}
\caption{A loop diagram, in which four external particles exchange a
  pair of virtual gluons. Once the momentum $k$ is specified, this
  fixes all other momenta as shown.}
\label{fig:loopdiag}
\end{center}
\end{figure}

The diagrams of figure~\ref{fig:Feynman} are the simplest ones we can
draw that connect all four external particles. For a given number of
external lines, we can also add more vertices by including more
virtual particles. An example is shown in figure~\ref{fig:loopdiag},
in which the two incoming gluons exchange a pair of virtual gluons.
If loops are present, the external momenta are no longer sufficient to
determine the momenta of all internal lines. We must fix a momentum
$k$ (we can choose which one), after which all the other momenta are
determined, as shown in the figure. You can easily convince yourself
that if $L$ loops are present, one needs to specify the momenta of $L$
internal lines in this way. These momenta are called {\it loop
  momenta}, and diagrams with no loops (such as those of
figure~\ref{fig:Feynman}) are called {\it tree-level diagrams}, as the
corresponding graphs constitute trees in graph theory parlance. 

Feynman diagrams gives us a convenient way to visualise how particle
interactions happen in quantum field theory. However, they are much
more than this. It turns out that there is a precise set of so-called
{\it Feynman rules}, that convert each diagram into a mathematical
contribution to a complex number called the {\it scattering amplitude}
${\cal A}$, where $|{\cal A}|^2$ is related to the probability for the
interaction to occur. In order to state the rules, one must break the
gauge invariance of the theory. This is called {\it choosing a gauge},
and different Feynman rules exist for different gauges. However, upon
adding together the contributions from all possible diagrams, the
resulting scattering amplitude becomes gauge-invariant, as it should
be given that this is a symmetry of the theory. A common gauge choice
for Yang-Mills theory is the {\it Feynman gauge}, which says that each
internal line of 4-momentum $q^\mu$ is associated with a factor
\begin{equation}
D^{ab}_{\mu\nu}=-\frac{i\delta^{ab}\eta_{\mu\nu}}{q^2},
\label{prop}
\end{equation}
where $(a,b)$ and $(\mu,\nu)$ are respectively the colour and Lorentz
indices associated with each end of the line. Furthermore,
$\delta^{ab}$ and $\eta_{\mu\nu}={\rm diag}(1,-1,-1,-1)$ are the
Kronecker delta symbol and the metric tensor for Minkowski space. We
see that every internal line involves an inverse power of the squared
4-momentum flowing through the line.

Each three-gluon vertex is associated with a factor
\begin{align}
V^{a_1 a_2 a_3}_{\mu_1\nu_1\nu_2}(p_1,p_2,p_3)=-g
f^{abc}&\left[(p_1-p_2)_{\mu_3}\eta_{\mu_1\mu_2}
+(p_2-p_3)_{\mu_1}\eta_{\mu_2\mu_3}\right.\notag\\
&\left.\quad+(p_3-p_1)_{\mu_2}\eta_{\mu_3\mu_1}\right].
\label{3gv}
\end{align}
Here $g$ is a {\it coupling constant} representing the strength of the
force between interacting gluons (in the full theory of QCD, it also
describes the interaction strength between gluons and quarks).  Each
gluon of momentum $p_i$ entering the vertex has an associated Lorentz
index $\mu_i$ and colour index $a_i$. Furthermore, $f^{a_1 a_2 a_3}$
is a known number which depends on the colour charges of the three
gluons~\footnote{More specifically, for any given gauge group, the
  quantities $\{f^{a_1 a_2 a_3}\}$ are the structure constants of the
  associated Lie algebra.}. It is antisymmetric under the interchange
of any two indices:
\begin{equation}
f^{abc}=f^{bca}=f^{cab}=-f^{acb}=-f^{cba}=-f^{bac},
\label{antisym}
\end{equation}
and satisfies the {\it Jacobi identity}
\begin{equation}
f^{abe}f^{ecd}+f^{ace}f^{edb}+f^{ade}f^{ebc}=0,
\label{Jacobi}
\end{equation}
where summation over repeated indices is implied. With each incoming
gluon, we must associate a 4-vector $\epsilon_\mu(p_i)$ (or its
complex conjugate if the gluon is outgoing) describing its
polarisation. The factors associated with each internal line, vertex
and external particle are then simply multiplied together for each
Feynman diagram, with sums over all repeated Lorentz and colour
indices. There is also a rule for the four gluon vertex appearing in
figure~\ref{fig:Feynman}(d). However, it turns out that one can always
choose to rewrite such diagrams to involve sums of products of
three-gluon vertices, so we will not need the four-vertex for the
following discussion.

For loop diagrams, an additional rule is needed. Given that each loop
has an associated loop momentum $k^\mu$, we need to sum over all the
possible values that this momentum can have. This takes the form of an
integral over all 4 components. Finally, for all diagrams, one must
divide by the number of symmetry transformations $S_i$ that leave the
graph invariant. From this and the above rules (including the
conservation of 4-momentum at all vertices), we can write a very
general schematic formula for any scattering amplitude in Yang-Mills
theory, involving an arbitary number $m$ of external particles, and
involving $L$ loops~\footnote{To be fully general, we present this
  formula in $D$ spacetime dimensions, where our apparent spacetime has
  $D=4$.}:
\begin{equation}
{\cal A}_m^{(L)}=i^Lg^{m-2+2L}\sum_{i}\int\prod_{l=1}^L\frac{d^Dk_l}
{(2\pi)^D}\frac{1}{S_i}\frac{n_i\,c_i}{\prod_{\alpha_i}p_{\alpha_i}^2}.
\label{ampform}
\end{equation}
The sum is over all cubic graphs $i$ (where it is assumed all
four-vertices have been eliminated). The overall power of the coupling
constant $g$ is equal to the number of cubic vertices. For each graph,
there is an integral over the loop momenta ($L$ of them), and a
denominator involving the product of all momenta associated with
internal lines. The numerator for each graph will contain two
contributions. Firstly, there is a part depending on the colour
charges of the gluons. From eq.~(\ref{3gv}), this is simply given by
the product of the factors $f^{a_1a_2a_3}$ at each vertex. For
example, the colour factors of the diagrams in
figure~\ref{fig:Feynman}(a)--(c) are given by~\footnote{To reproduce
  these expressions, one must assign the colour indices clockwise at
  each vertex. We have labelled colour indices for the external and
  internal gluon for all graphs as in figure~\ref{fig:Feynman}(a).}
\begin{equation}
c_a=f^{ace}f^{edb},\quad c_b=f^{aeb} f^{ecd},\quad c_c=f^{ade}f^{ecb}.
\label{colfacs}
\end{equation}
Secondly, there is a kinematic factor $n_i$, depending on the external
and loop momenta, as well as the polarisation vectors for the external
particles. Note that individual terms in eq.~(\ref{ampform}) do not
necessarily come from individual Feynman diagrams: for example, the
four-gluon vertex in the latter is rewritten so as to contribute to
multiple terms in the sum. Calculating the amplitude then amounts to
determining the kinematic numerators $\{n_i\}$ for each graph. They
are not unique - their explicit form will depend on the choice of
gauge, as well as other freedoms that we will see later. However, the
scattering amplitude is unique for a given theory, so that any
ambiguities must cancel in the sum in eq.~(\ref{ampform}).

We have so far described scattering amplitudes in a non-abelian gauge
theory. However, we can also talk about them in quantum gravity, which
means identifying a ``graviton field'' that is analogous to the field
$A_\mu^a$ describing the gluon in a gauge theory, and which carries
the force of gravity. We have already seen above that General
Relativity associates gravity with the structure of space-time
itself. We can represent this mathematically as follows. Consider two
events in spacetime, separated by an infinitesimal distance
\begin{equation}
dx^\mu=(dt,d\vec{x}),
\label{dxmudef}
\end{equation}
where $dt$ and $d\vec{x}$ are the temporal and (vector) spatial
separation respectively. In the absence of gravity, we can associate a
Lorentz-invariant spacetime distance
\begin{equation}
ds^2=\eta_{\mu\nu}dx^\mu dx^\nu=(dt)^2-(d\vec{x})\cdot(d\vec{x}),
\label{ds2def}
\end{equation}
where repeated indices are summed over, and $\eta_{\mu\nu}={\rm
  diag}(1,-1,-1,-1)$ is the metric tensor of flat space. When gravity
is present, this equation becomes modified to
\begin{equation}
ds^2=g_{\mu\nu}(t,\vec{x})dx^\mu dx^\nu.
\label{ds2def2}
\end{equation}
That is, the metric tensor becomes a function of spacetime position,
so that the distance measure changes as we move throughout the
space. The space is thus warped, and the form that $g_{\mu\nu}$ takes
is dictated by the distribution of mass and energy throughout the
spacetime. The precise relationship is given by {\it Einstein's field
  equations}, which take the form of a differential equation relating
second derivatives of the metric tensor (sensitive to the curvature of
spacetime) to an {\it energy-momentum tensor} describing the matter
distribution.

For weak gravitational fields, we know that the metric tensor must
reduce to its flat space counterpart. This allows to write
\begin{equation}
g_{\mu\nu}(t,\vec{x})=\eta_{\mu\nu}+\kappa h_{\mu\nu}(t,\vec{x})
\label{hdef}
\end{equation}
for some number $\kappa$, such that $h^{\mu\nu}$ is a field
representing the presence of gravity i.e. the deviation from flat
space. Substituting this into the Einstein equations, we may fix
$\kappa$ by requiring consistency with Newton's theory of gravitation
for sufficiently weak fields, which leads to $\kappa=\sqrt{32\pi
  G_N}$, where $G_N$ is Newton's constant. The number $\kappa$ thus
represents the strength of the gravitational force i.e. it is the
analogue of the coupling constant $g$ in a non-abelian gauge theory.

Substituting eq.~(\ref{hdef}) into the equations of General Relativity
leads to an equation for the field $h_{\mu\nu}(x)$, whose solutions
include the {\it gravitational waves} recently discovered by
LIGO~\cite{Abbott:2016blz}. The equation is highly nonlinear, where
the nonlinear terms correspond to interactions between the field
$h_{\mu\nu}$ and itself, in much the same way that the gluon can
interact with itself in gauge theory. Similarly to the latter case, we
can attempt to quantise the theory, such that gravitational waves come
in discrete packets called {\it gravitons}. We can then consider
scattering amplitudes for gravitons, which will be given by similar
Feynman diagrams to those we have already seen for gluons. However,
the Feynman rules will now be different, as they will be governed by
the equations of General Relativity rather than Yang-Mills theory.

This procedure was first carried out in earnest in the
1960s~\cite{Feynman:1963ax,DeWitt:1967yk,DeWitt:1967ub,DeWitt:1967uc}. As
for non-abelian gauge theories, the Feynman rules are not unique:
given that the graviton $h_{\mu\nu}$ has two free spacetime indices,
the rules will depend on which coordinate system is chosen, such that
this ambiguity cancels out in the final scattering amplitude. The fact
that final results must be independent of the coordinate system is
known as {\it diffeomorphism invariance}, and it is possible to
understand this as a type of gauge symmetry acting on the graviton
field $h_{\mu\nu}$. Fixing a coordinate system for the Feynman rules
is then referred to as {\it choosing a gauge}, by analogy with the
Yang-Mills case. In the commonly chosen {\it De Donder gauge}, the
three-graviton vertex has well over a hundred individual terms, in
stark contrast to the three-gluon vertex in Yang-Mills theory, which
has only six terms, as can be seen in eq.~(\ref{3gv}). Vertices
involving four or more gravitons are even more complicated, such that
calculations in quantum gravity are stupendously unwieldy, rapidly
becoming intractable as either the number of loops in Feynman
diagrams, or the number of external particles, is increased. The use
of powerful computers can help, but available computing power still
limits which calculations are possible. As discussed above, results
obtained in pure General Relativity indicate that this theory contains
divergences at higher loop orders which cannot be removed by simply
redefining the parameters of the theory (i.e. the theory is
non-renormalisable). It is possible that adding extra matter content
or symmetries to the theory may help, but investigation of this
question is clearly hampered by the difficult nature of amplitude
calculations.

Like Yang-Mills theory, individual Feynman diagrams in gravity are not
gauge-invariant by themselves, such that the symmetry is only restored
upon combining all diagrams to form the full amplitude. Remarkably,
the final results are often much simpler (exceedingly so in gravity)
than the individual diagrams would suggest. Literally thousands of
terms {\it cancel} in the sum over diagrams, suggesting that much of
the complexity in Feynman diagram calculations is associated with
gauge-dependent artifacts, that are physically irrelevant, and thus
ultimately absent in the final results.

The above considerations have led to the development of alternative
methods for calculating amplitudes, that don't necessarily use Feynman
diagrams as a starting point (see
e.g. refs.~\cite{Elvang:2013cua,Cheung:2017pzi} for pedagogical
reviews). The picture that is emerging is that gravity amplitudes, if
thought about in the right way, possess a simplicity and elegance that
is almost entirely hidden using traditional calculational
methods. Furthermore, there are deep and profound similarities between
gravity amplitudes and their counterparts in non-abelian gauge
theories, that may have far-reaching consequences.

\section{BCJ duality and the double copy}
\label{sec:BCJdouble}

\begin{figure}
\begin{center}
\scalebox{0.8}{\includegraphics{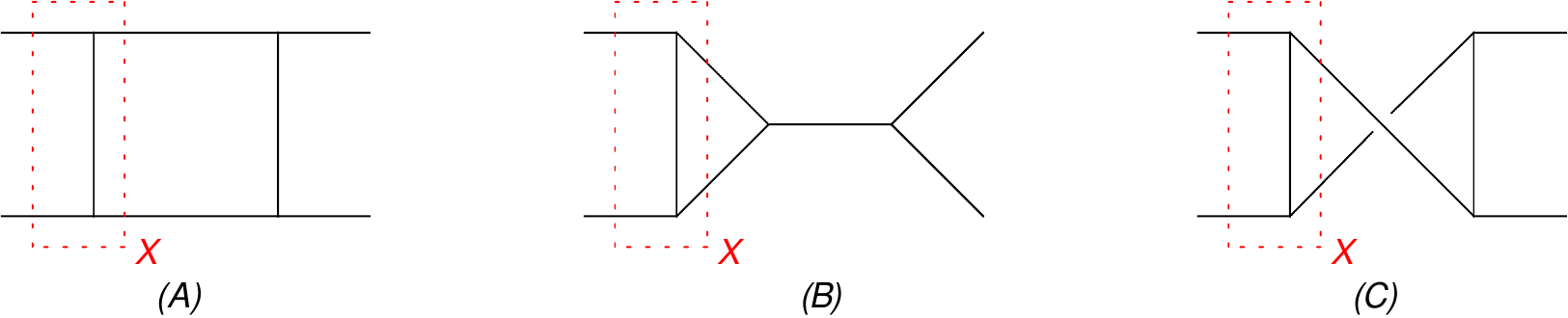}}
\caption{Three Feynman diagrams occuring in Yang-Mills theory at
  one-loop order, where gluons have been shown as solid lines for
  brevity. The highlighted region $X$ is the same in each diagram, and
  thus contributes a common colour factor.}
\label{fig:loopdiags}
\end{center}
\end{figure}

In eq.~(\ref{ampform}) we saw that a scattering amplitude in a
non-abelian gauge theory can be written in a very general form, as a
sum over cubic graphs. Each of these has a colour factor obtained by
dressing each vertex with a factor $f^{abc}$, where $\{a,b,c\}$ denote
the colour indices of the gluons entering the vertex. Furthermore,
these factors satisfy the Jacobi identity of eq.~(\ref{Jacobi}), which
allows us to relate the colour factors of different Feynman
diagrams. As an example, the results of eq.~(\ref{antisym},
\ref{Jacobi}, \ref{colfacs}) imply that the colour factors of the
diagrams in figure~\ref{fig:Feynman}(a)--(c) are related
by
\begin{equation}
c_a-c_b-c_c=0.
\label{colrel}
\end{equation}
Similar relations apply between more complicated diagrams, involving
more external particles, or loops. An example is shown in
figure~\ref{fig:loopdiags}, which shows three diagrams arising at
one-loop order. Each diagram consists of a part $X$ common to all
three, which will give rise to the same contribution $c_X$ to each
colour factor. There is then a part which looks like one of the
diagrams of figure~\ref{fig:Feynman}(a)--(c). We can then write the
colour factor of each diagram as 
\begin{equation}
c_A=c_X c_a,\quad c_B=c_X c_b,\quad c_C=c_X c_c,
\label{colX}
\end{equation}
such that eq.~(\ref{colrel}) implies
\begin{equation}
c_A-c_B-c_C=0.
\label{colrel2}
\end{equation}
Generalising this, we find that at any number of loops $L$, and for
any number of external particles $m$, we can always arrange diagrams
into overlapping sets of three, each obeying a {\it Jacobi relation}
such as those of eqs.~(\ref{colrel}, \ref{colrel2}). These relations
are a direct consequence of the non-abelian gauge symmetry of the
theory, which gives rise to the properties of eq.~(\ref{antisym},
\ref{Jacobi}).

In 2008, a remarkable new property of tree-level ($L=0$) gauge theory
amplitudes was discovered: whenever a given set of three diagrams
$\{i,j,k\}$ obeys a colour Jacobi relation, the corresponding
kinematic numerators (as functions of momenta) can be chosen to obey a
similar relation:
\begin{equation}
c_i\pm c_j \pm c_k=0\quad \Rightarrow \quad
n_i\pm n_j \pm n_k=0.
\label{BCJdef}
\end{equation}
Furthermore, the numerators $\{n_i\}$ can be chosen to obey similar
antisymmetry conditions to eq.~(\ref{antisym}), but where kinematic
rather than colour degrees of freedom are interchanged. This property
is known as {\it BCJ duality}, and had not been previously
appreciated, for understandable reasons. Recall that the kinematic
numerators are not unique, but depend e.g. on the gauge chosen for
calculating Feynman diagrams. In arbitrary gauges (including those
most commonly used for calculations), the numerators {\it will not}
obey eq.~(\ref{BCJdef}). However, it is always possible to shift the
numerators according to
\begin{equation}
n_i\rightarrow n_i+\Delta_i,
\label{gengauge}
\end{equation}
where $\Delta_i$ is a function of momenta, such that BCJ duality is
satisfied. This is called a {\it generalised gauge transformation}, as
it combines a gauge transformation with other possible operations,
such as a redefinition of the gluon field. Substituting
eq.~(\ref{gengauge}) into eq.~(\ref{ampform}), a generalised gauge
transformation is permitted provided the shifts $\{\Delta_i\}$ satisfy
\begin{equation}
\sum_i \frac{1}{S_i}\frac{\Delta_i\,c_i}{\prod_{\alpha_i}p_{\alpha_i}^2}=0,
\label{gengauge2}
\end{equation}
which restricts the possible form of each $\Delta_i$, but still allows
a great deal of freedom. 

BCJ duality is proven to be possible at tree-level and has been
conjectured to hold at all loop orders, due to highly nontrivial
evidence. For example, we have here described the simplest theory in
which BCJ duality applies, namely pure Yang-Mills theory. However,
supersymmetric generalisations of this theory exist, in which
additional fermionic and scalar matter is added, whose couplings are
such that the equations of the theory are invariant under
interchanging bosonic and fermionic degrees of freedom. Loop
calculations become easier in such theories, and in a particular
example (${\cal N}=4$ Super-Yang-Mills theory), BCJ duality has been
observed up to four-loop order~\cite{Bern:2010ue,Bern:2012uf}. It has
also been seen at one and two-loop order even if supersymmetry is not
present~\cite{Bern:2013yya}, and thus BCJ duality in some form is
certainly a genuine quantum property of nature. 

We saw that the colour Jacobi relations are a manifestation of
non-abelian gauge symmetry. The fact that the kinematic numerators
satisfy similar relations implies that there must also be a
(different) symmetry underlying them. What this symmetry might be
remains unknown. It is also unclear whether such a symmetry underlies
the full equations defining the theory, or is merely an accidental
property of scattering amplitudes. Intriguing hints, however, come
from {\it self-dual Yang-Mills theory}, which is obtained from
Yang-Mills theory by keeping only a single, particular polarisation
state of the gluon. In this case, the kinematic symmetry has been
associated with a certain area-preserving diffeomorphism
group~\cite{Monteiro:2011pc}. If more general results for full
Yang-Mills and related theories can be elucidated, it would constitute
a hitherto undiscovered structure in non-abelian gauge theory, that
may provide deep insights into why such theories are so fundamental in
nature. Even without this, imposing BCJ duality may be a useful tool
for calculating new scattering amplitudes (see
e.g. ref.~\cite{Badger:2015lda}).

Once the numerators in eq.~(\ref{ampform}) have been chosen to make
BCJ duality manifest, a second remarkable conjecture links gauge
theories with gravity, as follows. First, one may replace the gauge
theory coupling $g$ with its gravitational counterpart, according to
\begin{equation}
g\rightarrow \frac{\kappa}{2}.
\label{greplace}
\end{equation}
Next, one can strip off the colour factors $c_i$ for each cubic graph,
and replace them with a second set $\{\tilde{n}_i\}$ of kinematic
numerators. The tilde on these quantities indicates that they do not
necessarily come from the same gauge theory as the $\{n_i\}$. Upon
making these modifications, one arrives at the quantity~\footnote{The
  additional factor of $i$ in eq.~(\ref{ampform2}) relative to
  eq.~(\ref{ampform}) is conventional, and will cancel in the squared
  amplitude.}
\begin{equation}
{\cal M}_m^{(L)}=i^{L+1}\left(\frac{\kappa}{2}\right)^{m-2+2L}
\sum_{i}\int\prod_{l=1}^L\frac{d^Dp_l}
{(2\pi)^D}\frac{1}{S_i}\frac{n_i\,\tilde{n}_i}{\prod_{\alpha_i}p_{\alpha_i}^2}.
\label{ampform2}
\end{equation}
The {\it double copy}
conjecture~\cite{Bern:2010ue,Bern:2010yg}~\footnote{It can be proven
  that the double copy holds provided BCJ duality does at higher loop
  orders, as explained in ref.~\cite{Bern:2010yg}.} states that this
formula corresponds to a gravitational scattering amplitude with $m$
external particles and $L$ loops. Even without knowing any physics at
all, it is clear from sight alone that eq.~(\ref{ampform2}) is
extremely similar to its gauge theory counterpart,
eq.~(\ref{ampform})!  The name ``double copy'' refers to the fact that
two gauge theory numerators occur in the gravity formula. This is
sometimes also expressed by saying that gravity is the ``square'' of
gauge theory, although we do not mean that one literally squares the
gauge theory amplitude to obtain the gravity result: each term is
copied separately, and denominator factors are left untouched.

It is difficult to exaggerate how remarkable eq.~(\ref{ampform2})
is. The traditional calculation of gravity amplitudes, as discussed
above, is extraordinarily complicated, involving a forbidding quagmire
of algebraic complexity. Eq.~(\ref{ampform2}), however, suggests that
the basic structure of gravity amplitudes is essentially identical to
those of gauge theory, despite the fact that the Feynman rules are
orders of magnitude more complicated! 

We have not yet stated which gravity theory the amplitude of
eq.~(\ref{ampform2}) resides in, and indeed the answer depends on
which two gauge theories are chosen for the two sets of kinematic
numerators. The simplest form of the double copy consists of pure
Yang-Mills theory being copied with itself. This leads to General
Relativity coupled to an additional scalar particle (known as the {\it
  dilaton}), and an antisymmetric tensor field. In four dimensions,
the extra particles constitute two degrees of freedom. Combined with
the two polarisation states of the graviton, this gives four degrees
of freedom in total. This matches up with the fact that we have
``copied'' two gauge theories with 2 degrees of freedom each (the two
polarisation states of the gluon). More complicated examples of the
double copy involve e.g. supersymmetric generalisations of Yang-Mills
theory. If we have ${\cal N}=N$ supersymmetries (relations between
bosons and fermions) in the first gauge theory, and ${\cal N}=M$ in
the second, the number of supersymmetries in the resulting gravity
theory is ${\cal N}=(N+M)$. Again, the number of degrees of freedom
(i.e. the number of individual polarisation states) on both sides of
the correspondence have to match up.

How do we know that eq.~(\ref{ampform2}) corresponds to a gravity
amplitude (possibly with additional matter)? First of all, we can
compare known gravity amplitudes with the result of
eq.~(\ref{ampform2}), and verify the agreement directly. Alternative
methods exist for determining whether a previously unknown amplitude
is indeed physically correct (see e.g.~\cite{Elvang:2013cua} for a
review). Further evidence for the correctness of the double copy
conjecture comes from examining particle scattering in certain special
energy regimes. In the limits when particles have a very high centre
of mass
energy~\cite{Saotome:2012vy,Vera:2012ds,Johansson:2013nsa,Johansson:2013aca,Melville:2013qca,Luna:2016idw},
or when very fast-moving particles exchange very low energy gravitons
or
gluons~\cite{Weinberg:1965nx,Naculich:2011ry,Laenen:2008gt,White:2011yy,Akhoury:2011kq,Beneke:2012xa,Oxburgh:2012zr},
the double copy can be verified for any number of loops $L$. Beyond
this, the lack of a formal proof of the double copy is related to the
lack of a formal proof of BCJ duality beyond tree level.

The double copy consists of stripping off the colour factors from a
gauge theory amplitude, and replacing them with extra kinematic
factors. We could also do the opposite, namely replacing the kinematic
factors $\{n_i\}$ with a second set of colour factors
$\{\tilde{c}_i\}$. This is called the {\it zeroth copy}, and yields
the quantity
\begin{equation}
{\cal T}_m^{(L)}=i^Ly^{m-2+2L}\sum_{i}\int\prod_{l=1}^L\frac{d^Dp_l}
{(2\pi)^D}\frac{1}{S_i}\frac{c_i\,\tilde{c}_i}{\prod_{\alpha_i}p_{\alpha_i}^2},
\label{ampform3}
\end{equation}
where we have also replaced the coupling $g\rightarrow y$. We can then
ask if eq.~(\ref{ampform3}) corresponds to a scattering amplitude in
some theory, and the answer is indeed yes. The relevant theory is
called {\it biadjoint scalar theory}, and consists of a scalar field
$\Phi^{aa'}$ that carries two different types of colour charge,
labelled by $a$ and $a'$. The field interacts with itself, where $y$
describes the strength of this interaction. Although this is not a
physical theory itself, there is mounting evidence that dynamics in
this theory is inherited, via the double copy, by gauge theory and
gravity. The network of theories which are related by similar
relationships continues to
widen~\cite{Johansson:2017srf,Anastasiou:2017nsz,Borsten:2017jpt,Cheung:2017ems}.

\section{Insights from string theory}
\label{sec:string}

\begin{figure}
\begin{center}
\scalebox{0.5}{\includegraphics{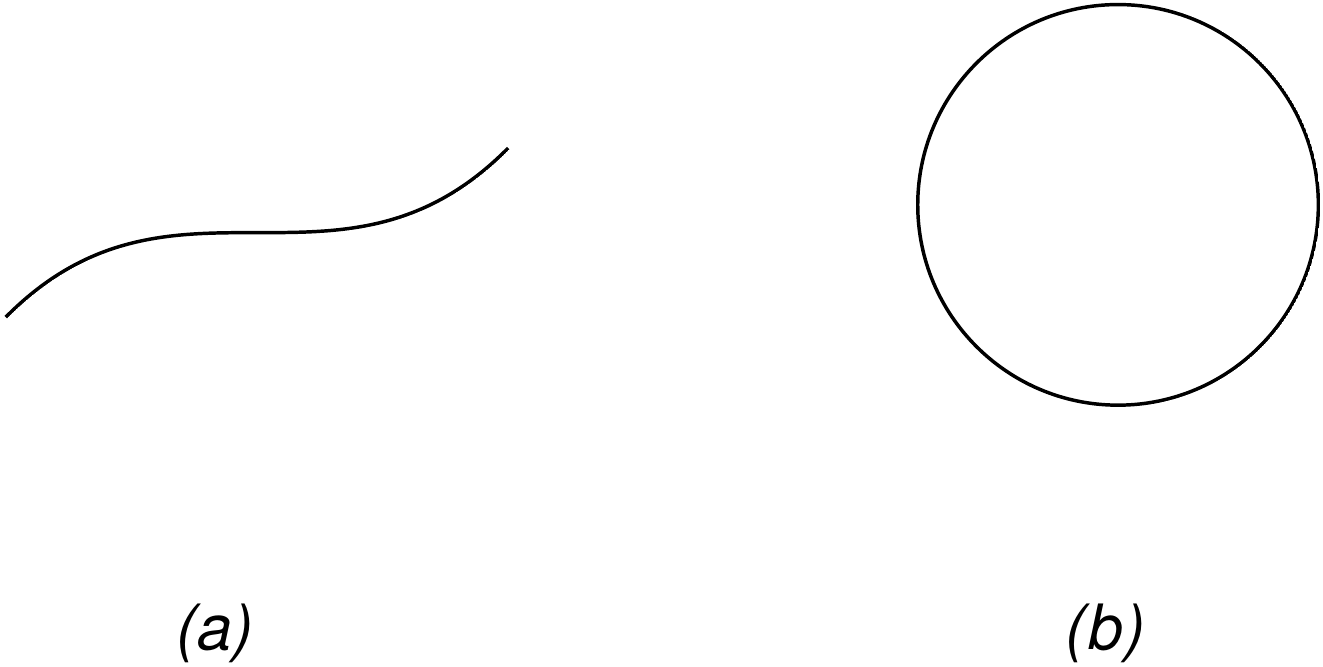}}
\caption{(a) An open string; (b) a closed string.}
\label{fig:strings}
\end{center}
\end{figure}

\begin{figure}
\begin{center}
\scalebox{0.5}{\includegraphics{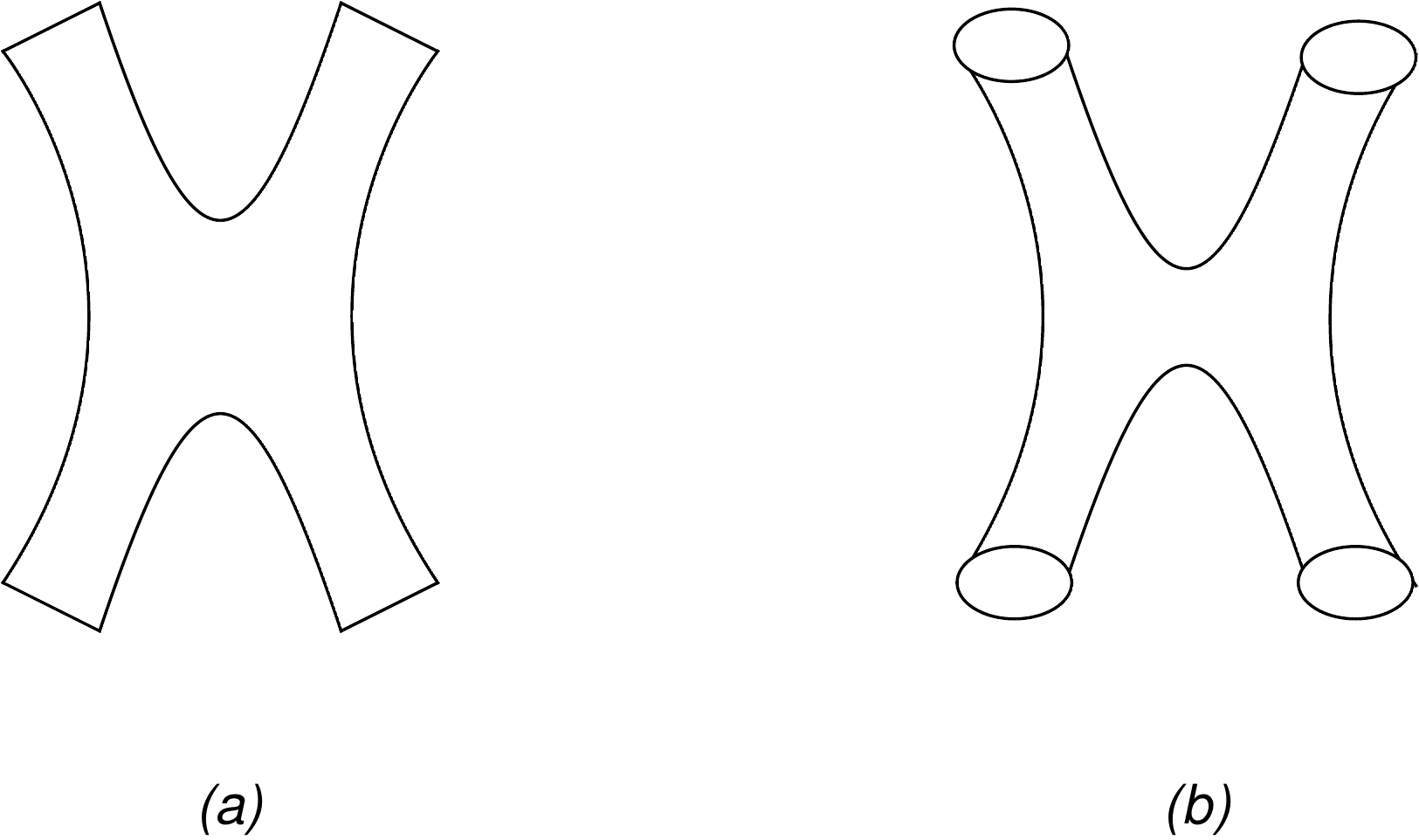}}
\caption{A string moving through spacetime traces out a
  worldsheet. These combine into Feynman diagrams, shown here for (a)
  four open strings interacting; (b) four closed strings.}
\label{fig:strings2}
\end{center}
\end{figure}

For tree-level diagrams (i.e. with no loops), the double copy turns
out to be equivalent to a previously discovered relationship between
scattering amplitudes in gauge and gravity theories, which can be
derived from {\it string theory}. The latter is a candidate for a
possible {\it theory of everything}, in which all the Standard Model
forces and gravity can be combined into a single consistent
theoretical description. The basic idea of string theory is that what
we see as particles at currently accessible energies are in fact
string-like at very small distance scales (from the uncertainty
principle, this is equivalent to very high energies)~\footnote{String
  theory also necessarily contains higher-dimensional objects called
  {\it branes}, that we will not need to discuss here.}. There are two
types of string: {\it open} and {\it closed}, as shown in
figure~\ref{fig:strings}. Strings can interact with each other, such
that we can draw their interactions using Feynman diagrams. However,
whereas a particle traces out a {\it worldline} as it moves through
spacetime, a string traces out a two-dimensional {\it
  worldsheet}. Examples for open and closed strings are shown in
figure~\ref{fig:strings2}.
It is known how to calculate scattering amplitudes for strings, and in
the absence of loops, the {\it KLT relations} state that amplitudes
for closed strings can be written as sums of products of amplitudes
for open strings~\cite{Kawai:1985xq}. The low energy limit of string
theory is a quantum field theory, where open strings give rise to
particles like the gluon, and closed strings correspond to
gravitons. Thus, the KLT relations in string theory imply a set of
double copy relations between scattering amplitudes in non-abelian
gauge theories and gravity. Note in this argument that it is
completely irrelevant whether or not string theory is a genuine
physical theory of nature. Here, it merely acts as a mathematical
bridge between two types of theory that definitely do exist in the
real world. 

Unfortunately, the string theory argument is not straightforward to
generalise to diagrams with loops, so that whether or not it provides
a full justification for the double copy remains unclear. Recent
developments, however, may shed light on this issue. Firstly, it is
possible to write field theory scattering amplitudes in a form that
looks very like a string theory, using the so-called {\it CHY
  equations}~\cite{Cachazo:2013iea,Cachazo:2013hca,Cachazo:2013gna}. Indeed,
the same formula has been shown to be obtainable from a string theory
living in {\it ambitwistor
  space}~\cite{Mason:2013sva,Geyer:2014fka,Casali:2015vta}, which can
itself be obtained as a limit of string theory in conventional
spacetime~\cite{Casali:2016atr}. Loop corrections can be calculated in
this theory~\cite{Geyer:2015bja,Geyer:2015jch}, and the CHY equations
are written in such a way that the double copy between gauge and
gravity theories becomes particularly clear. Thus, this may indeed
provide a framework for exploring the underlying origin of the double
copy. In a different approach, ref.~\cite{Cheung:2016say} (following
earlier related work that predates the double copy~\cite{Bern:1999ji})
rewrites the equations defining General Relativity in a new form,
which aims to make its double copy structure more manifest.

\section{The classical double copy}
\label{sec:classical}

Up to now, our discussion of the double copy has been restricted
solely to scattering amplitudes, which are approximate solutions of
the equations of quantum field theory (i.e. they correspond to an
expansion in the coupling constant). It is natural to ponder whether
the double copy goes much deeper than this, in which case it
constitutes a very profound relationship between gauge and gravity
theories, providing a common way to think about two previously
orthogonal physical frameworks. One way to examine this is to look at
{\it exact} solutions of gauge and gravity theories. This is a
two-step procedure: (i) one must find a way to associate a gauge field
$A_\mu^a$ with a given exact graviton solution $h_{\mu\nu}$ of general
relativity; (ii) one must show that the double copy relationship thus
obtained is consistent with the BCJ double copy for
amplitudes~\cite{Bern:2010ue,Bern:2010yg}.

This idea was first explored in ref.~\cite{Monteiro:2014cda}, which
considered the special class of {\it Kerr-Schild} metric tensors in
General Relativity. These take the form of eq.~(\ref{hdef}), where the
graviton field is given by
\begin{equation}
h_{\mu\nu}=\phi(t,\vec{x})k_\mu k_\nu.
\label{hKS}
\end{equation}
Here $\phi(t,\vec{x})$ is a scalar field, and we see that the graviton
consists of an outer product of a 4-vector $k_\mu$ with itself. This
vector is not arbitrary, but must satisfy the following two
conditions:
\begin{equation}
k^2=k^\mu k_\mu=0,\quad k^\mu\partial_\mu k^\nu=0.
\label{kKS}
\end{equation}
Substituting the ansatz of eq.~(\ref{hKS}) into the Einstein equations
results in them becoming {\it linear}. They are then relatively easy
to solve for $\phi$ and $k_\mu$, where different solutions correspond
to different possible spacetimes. Furthermore, the resulting solutions
are {\it exact}, with no higher order corrections in $\kappa$.

The simple form of eq.~(\ref{hKS}) suggests how one can form a gauge
field corresponding to the gravity solution: given that the gauge
field should contain one spacetime index rather than two, we can strip
off one copy of the Kerr-Schild vector $k_\nu$, and replace it with an
arbitrary constant vector in the abstract colour space of the gauge
theory. The resulting gauge field then has the form
\begin{equation}
A_\mu^a=c^a \phi(t,\vec{x})k_\mu,
\label{AmuKS}
\end{equation}
and ref.~\cite{Monteiro:2014cda} proved that a gauge field thus
obtained from any stationary Kerr-Schild graviton (i.e. with no
explicit time dependence) automatically satisfies the equations of
Yang-Mills theory. These equations also become linear, so that the
solution is exact, with no higher order corrections. This fulfills the
first condition above, of finding an exact map between solutions of
gauge theory and gravity. However, the procedure appears ambiguous:
why should one strip off the vector $k_\mu$, and not, say, absorb some
(or all) of the field $\phi(t,\vec{x})$ whilst doing so? In other words,
what fixes the double copy procedure between the two theories? The
answer can be obtained by considering the {\it zeroth copy}. Repeating
the above procedure, one obtains a biadjoint scalar field
\begin{equation}
\Phi^{aa'}=c^a\,\tilde{c}^{a'} \phi(t,\vec{x}),
\label{PhiKS}
\end{equation}
where we have replaced the Kerr-Schild vector in the gauge theory
solution by a second colour vector. The resulting field turns out to
be an exact solution of the theory whose amplitudes are given by
eq.~(\ref{ampform3}). Furthermore, the fact that one does not modify
the scalar field $\phi(t,\vec{x})$ when double copying exact solutions
from gauge theory to gravity can be related directly to the fact that
one does not modify denominator factors when double copying the
amplitudes of eqs.~(\ref{ampform}, \ref{ampform2},
\ref{ampform3}). This fulfills the second condition above, of showing
that the double copy for exact solutions is consistent with the
similar procedure for amplitudes.

Although the class of stationary Kerr-Schild solutions is rather
special, it is infinitely large, and includes well-known exact
solutions of General Relativity (albeit some that are more
conventionally presented in alternative coordinate systems). One
example is the Schwarzschild black hole, a static, non-rotating black
hole that would result, for example, from a completely symmetric
collapsing star. The metric for this solution can be written in
the Kerr-Schild form of eq.~(\ref{hKS}), where
\begin{equation}
\phi=\frac{M}{4\pi r},\quad k^\mu=\left(1,\hat{e}_r\right),
\label{Schwarzschild}
\end{equation}
where $r$ is the radial distance from the origin (where the centre of
the black hole is), and $\hat{e}_r$ a unit 3-vector in the radial
direction. The resulting gauge field, after making a gauge
transformation and the coupling replacement of eq.~(\ref{greplace}),
is found to be~\footnote{The mass $M$ in eq.~(\ref{Schwarzschild}) is
  also replaced with a colour-charge dependent factor that does not
  appear in eq.~(\ref{Coulomb}), as described in
  ref.~\cite{Monteiro:2014cda}.}
\begin{equation}
A^{\mu\,a}=\frac{c^a}{4\pi r}\left(1,\vec{0}\right).
\label{Coulomb}
\end{equation}
This is the well-known {\it Coulomb solution}, corresponding to a
static point charge at the origin. Thus, the single copy or ``square
root'' of a Schwarzschild black hole (sourced by a point-like mass),
is a point-like colour charge! This is conceptually similar to the
amplitude story, where colour (charge) information in the gauge theory
gets replaced by kinematic (i.e. momentum) factors. 

\begin{figure}
  \begin{center}
    \scalebox{0.5}{\includegraphics{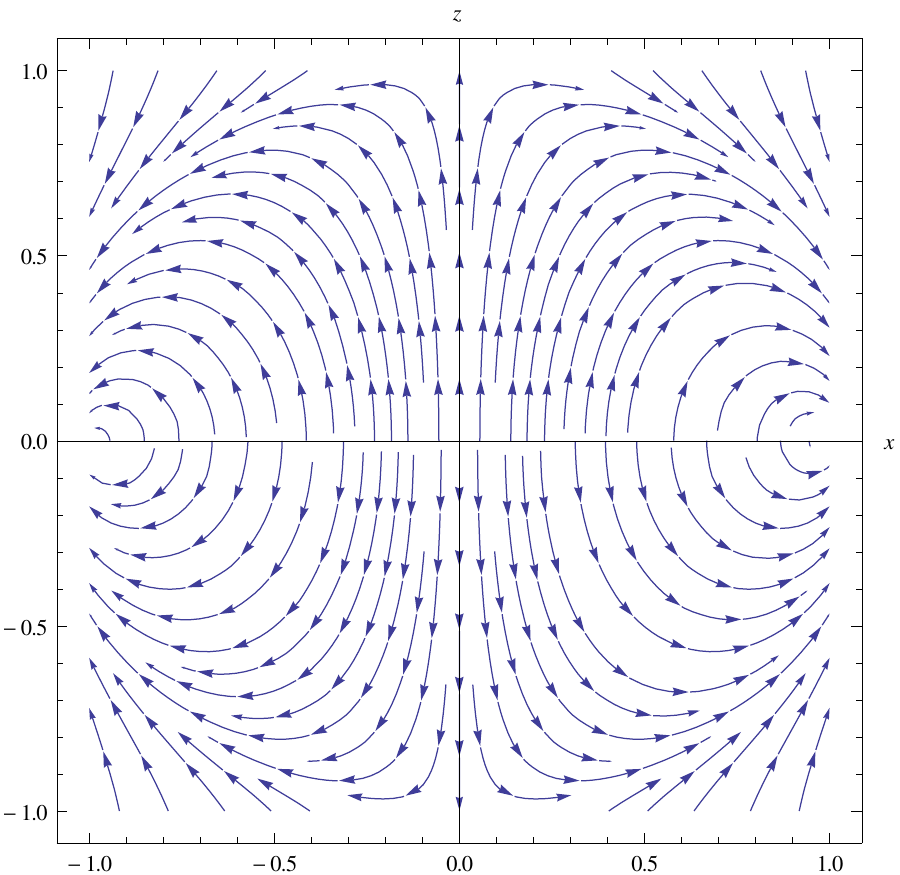}}
    \scalebox{0.5}{\includegraphics{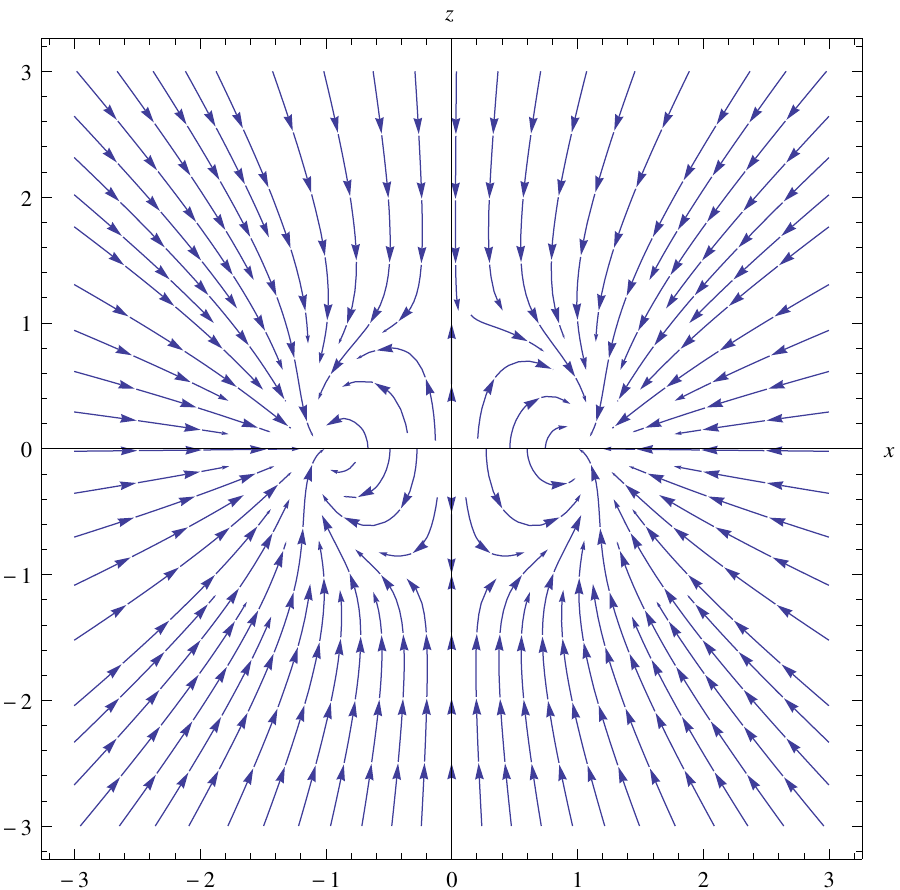}}
    \scalebox{0.5}{\includegraphics{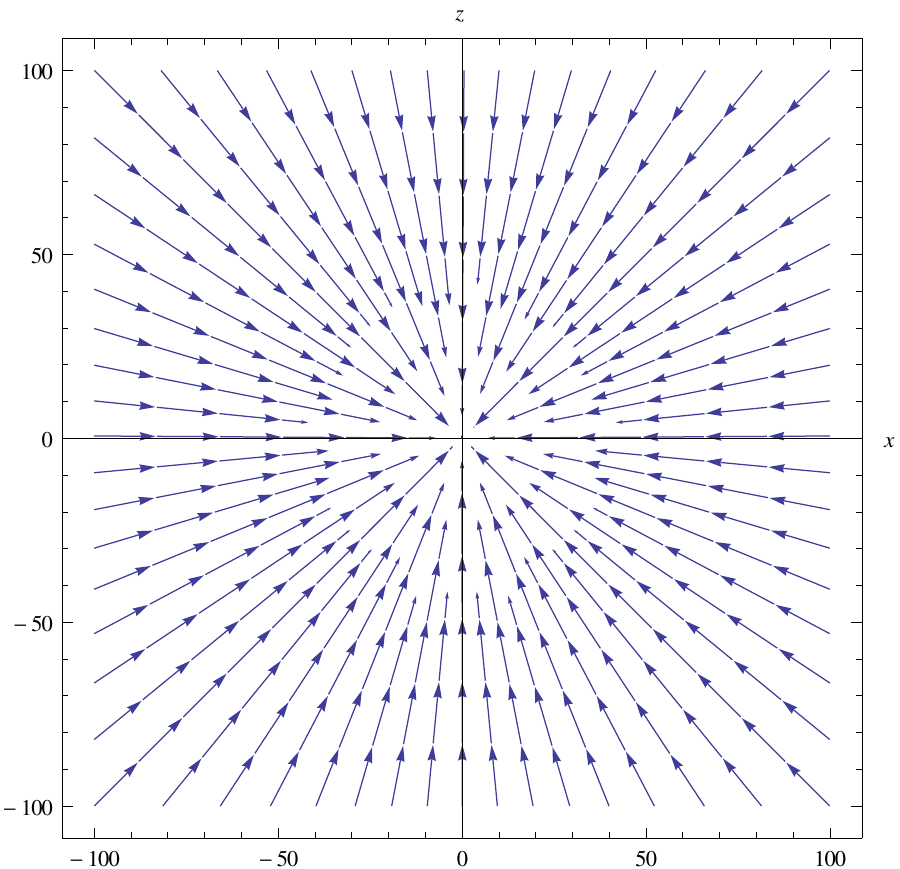}}
  \end{center}
\caption{The electric field generated by the single copy of the Kerr
  black hole, where the charged disk has unit radius and lies on the
  horizontal axis.}
\label{fig:Efield}
\end{figure}

\begin{figure}
  \begin{center}
    \scalebox{0.5}{\includegraphics{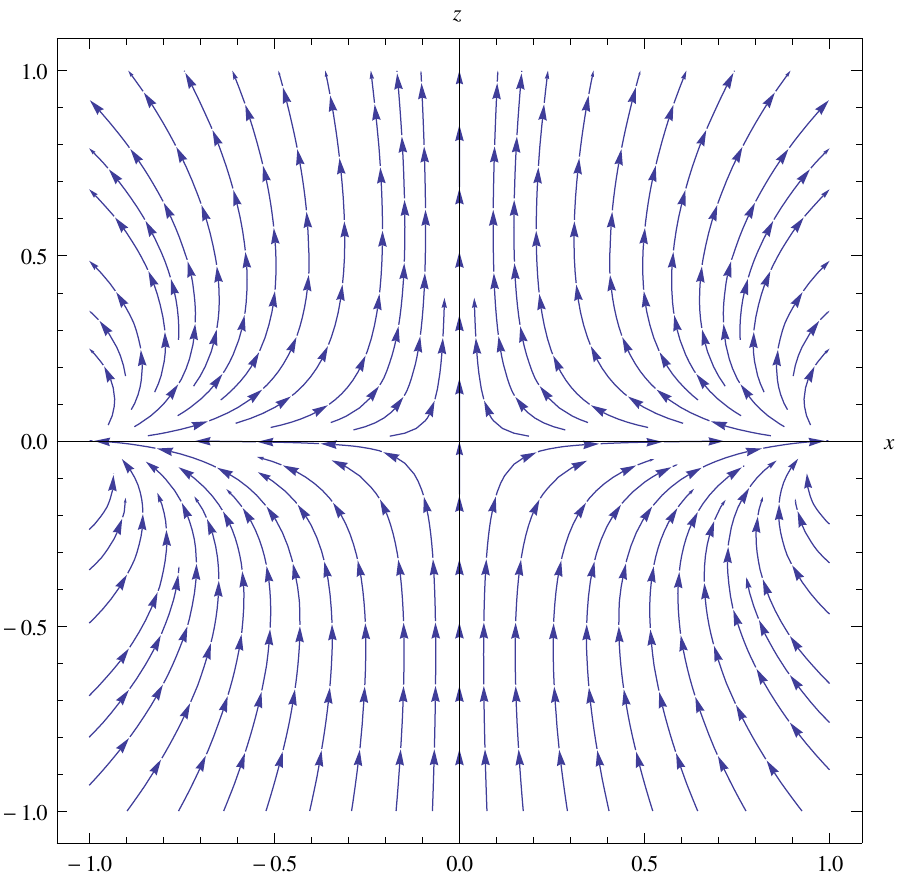}}
    \scalebox{0.5}{\includegraphics{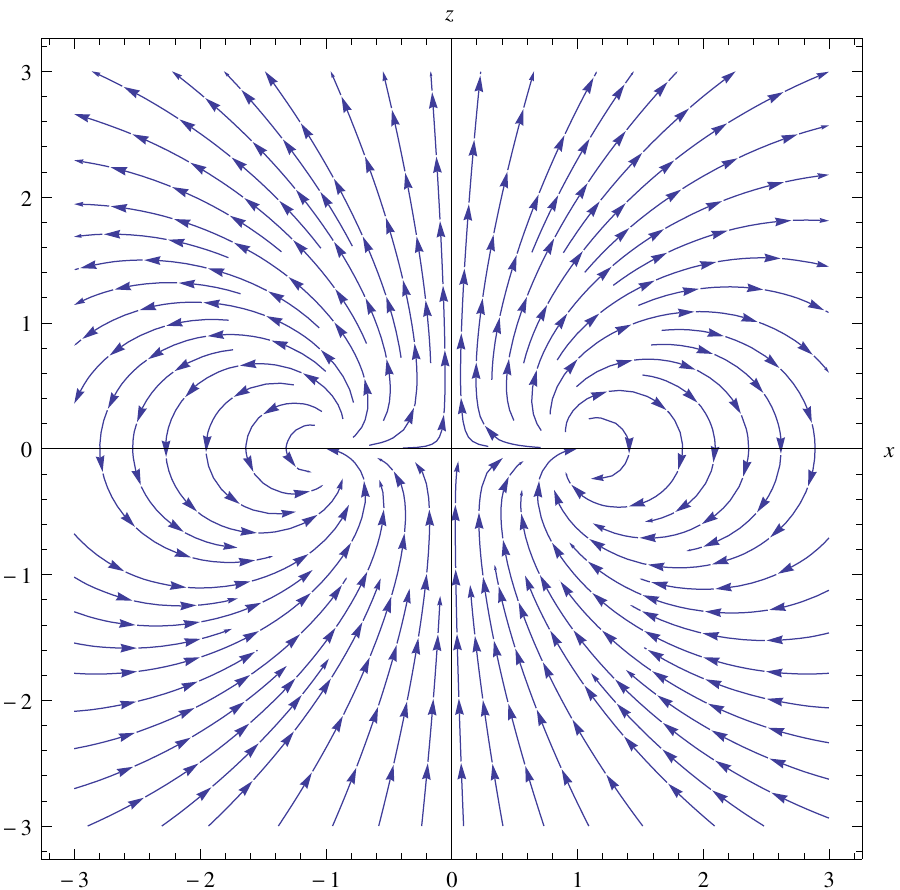}}
    \scalebox{0.5}{\includegraphics{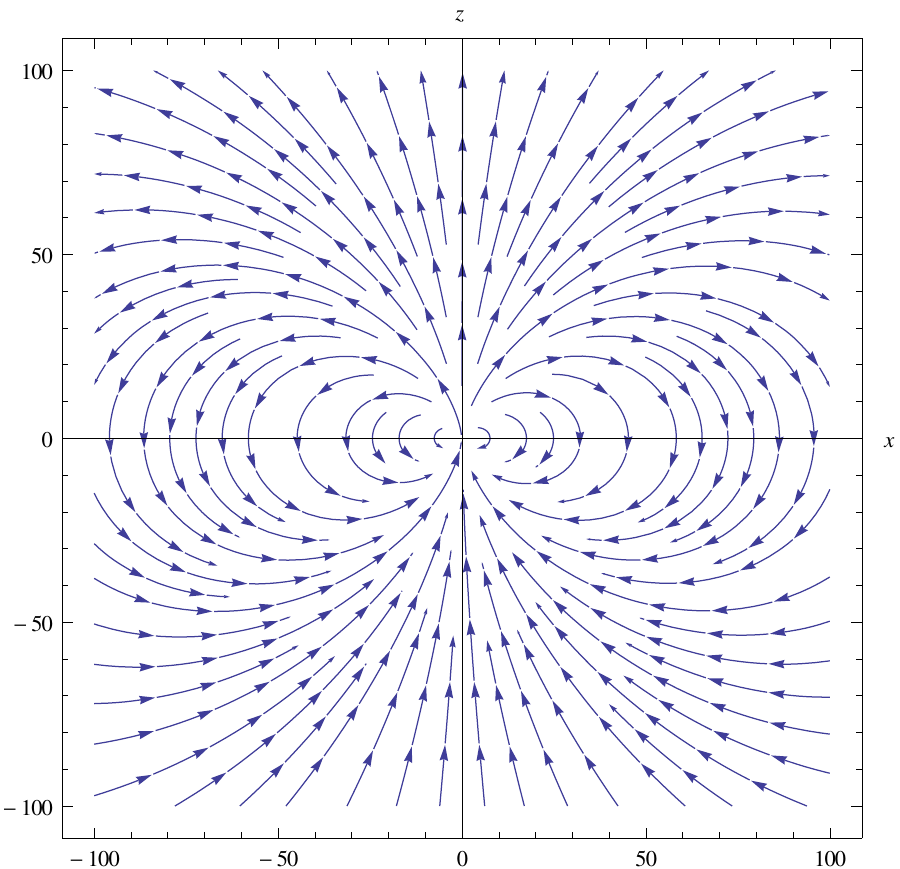}}
  \end{center}
\caption{The magnetic field generated by the single copy of the Kerr
  black hole, where the charged disk has unit radius and lies on the
  horizontal axis.}
\label{fig:Bfield}
\end{figure}

A more complicated black hole is the {\it Kerr black hole},
corresponding to a rotating disk of mass of finite size. The
corresponding gauge theory solution is found, as expected, to be a
rotating disk of charge, whose profile mirrors the mass profile in the
gravity theory~\cite{Monteiro:2014cda}. One can plot the electric and
magnetic fields in the gauge theory~\footnote{Given that the
  Yang-Mills equations linearise for this solution, they behave just
  like Maxwell's equations, so that we can talk about electric and
  magnetic fields in the same way.}, as we show in
figures~\ref{fig:Efield} and~\ref{fig:Bfield}. Both fields have
non-trivial structure at short distance, due to how the charge is
distributed on the disk. However, at large distances the electric
field looks like a point charge (i.e. points radially outwards) as it
should. Furthermore, the magnetic field is dipole-like, which is what
one expects for a coil of current. A rotating charge disk indeed looks
like a set of current coils, thus the results make sense.
Analogues of both the Schwarzschild and Kerr black holes exist in
higher dimensions~\cite{Tangherlini:1963bw,Myers:1986un}, and the
double copy works there too, which is consistent with the fact that
the BCJ copy for amplitudes is independent of the number of spacetime
dimensions $D$ in eqs.~(\ref{ampform}, \ref{ampform2},
\ref{ampform3}).

Whilst the fields double copy between gauge theory and gravity, the
same is not always true for the sources. For example, the
energy-momentum tensor that sources the Kerr metric is not a
straightforward double copy of the current corresponding to the
rotating charge disk in the gauge theory, but instead includes
additional radial pressure terms, which are needed to stablise the
mass disk in gravity~\cite{Monteiro:2014cda}. The nature of sources
has been further examined in ref.~\cite{Ridgway:2015fdl}.

More recent work has tried to extend the double copy for exact
solutions beyond the special case of Kerr-Schild. It is possible, for
example, to look at {\it multi-Kerr-Schild} solutions, in which
additional terms of the form of eq.~(\ref{hKS}) are added to the
metric, involving different scalar fields and 4-vectors. Such an
ansatz is no longer guaranteed to linearise the Einstein equations,
but a known case where this indeed happens is the {\it Taub-NUT
  metric}~\cite{Chong:2004hw}, originally derived (in a different
coordinate system) in refs.~\cite{Taub,NUT}. This solution has a
Schwarzschild-like mass at the origin, as well as a so-called {\it NUT
  charge} that gives rise to a rotational character of the
gravitational field at spatial infinity. The single copy was
constructed in ref.~\cite{Luna:2015paa} and consists of a {\it dyon}
having both electric and magnetic monopole charges, which map to the
mass and NUT charge in gravity respectively. Analogies between the
Taub-NUT solution and magnetic monopoles, at large radial distances,
have been made before. However, the double copy makes this
relationship exact, for arbitrary distances.

Another extension away from stationary Kerr-Schild metrics is to
consider time dependence. Arguably the simplest time-dependent
solutions are plane waves~\cite{Coleman:1977ps,Aichelburg:1970dh},
which are known to double copy between gauge theories and
gravity~\cite{Siegel:1999ew,Aichelburg:1970dh,Barnett:2014era}, but
which can also be cast in Kerr-Schild
language~\cite{Monteiro:2014cda}. Wave-like solutions have also been
recently used to construct a double copy with a curved-space
(non-Minkowski) background for the
graviton~\cite{Adamo:2017nia}. Another time-dependent system is that
of an arbitrarily accelerating particle, considered in
ref.~\cite{Luna:2016due}. This has a known Kerr-Schild description
(see e.g. ref.~\cite{Stephani:2003tm}), but has the unusual feature
that radiation appears as a source term in both the Yang-Mills and
gravity equations~\cite{Luna:2016due}. As explained in the latter
reference, this can itself be related to known amplitudes for the
emission of gluon and graviton radiation in the low energy limit,
strengthening the link between the Kerr-Schild and amplitude double
copies yet further.

Kerr-Schild solutions are useful in that they linearise the field
equations, and hence are exact solutions. One can also consider
classical fields that are not known exactly, but must be constructed
order-by-order in the coupling (analagous to how scattering amplitudes
are calculated in perturbation theory). Indeed, such fields are
routinely calculated in order to compare GR with astrophysical
measurements, including those involving gravitational waves. Even in
classical GR, these calculations are highly complex, and would be
greatly simplified by being able to perform simpler calculations in a
gauge theory, before doubly copying them to gravity. This has been
investigated recently in
refs.~\cite{Goldberger:2016iau,Goldberger:2017frp,Luna:2016hge} (see
also~\cite{Bjerrum-Bohr:2013bxa,Bjerrum-Bohr:2016hpa}). Further work
is needed, however, in order to remove the additional matter particles
that appear alongside gravity in the double copy, and also to push the
calculations to higher powers of the coupling.

An orthogonal body of work has looked at constructing gravity
solutions, with and without supersymmetry, using convolutions of gauge
fields~\cite{Anastasiou:2014qba,Anastasiou:2016csv,Anastasiou:2017nsz,Borsten:2017jpt}. This
approach has the advantage of being independent of the gauge chosen on
the gauge theory side, but is currently set up only to linear order.

Finally, all of the above examples of the double copy rely on
solutions of the linearised field equations, in biadjoint scalar,
gauge or gravity theories. If a true double copy relationship exists
between these theories, it should be true also for fully {\it
  nonlinear} solutions, involving e.g. inverse powers of the
coupling. Preliminary steps to investigate this have been taken in
ref.~\cite{White:2016jzc,DeSmet:2017rve}, which derived a number of
nonlinear classical solutions of biadjoint scalar theory, with the
hope of using these as building blocks for such solutions in gauge
theory. Indeed, an intriguing relationship was noted between these
solutions and {\it Wu-Yang monopoles}~\cite{Wu:1967vp} in Yang-Mills
theory.

\section{Conclusion}
\label{sec:conclude}

The double copy is a new and remarkable relationship between gauge and
gravity theories, which underly the four fundamental forces in
nature. Furthermore, more exotic theories (such as biadjoint scalar
theory) can be shown to obey double-copy like behaviour, such that the
dynamics of both gauge and gravity theories may be a lot simpler than
previously thought.

There are a number of both theoretical and practical applications of
the double copy. Firstly, it greatly streamlines calculations in both
classical and quantum gravity. This has renewed the investigation of
whether alternative field theories of gravity - such as ${\cal N}=8$
Supergravity - are well-behaved in the quantum
regime~\cite{Bern:2007hh,Bern:2009kd}. The calculation of
gravitational observables for use in astrophysics may potentially also
be drastically simplified, by essentially replacing GR with
double-copied Yang-Mills theory. Such techniques may also prove useful
for cosmology. The new way to think about gravity provided by the
double copy may give clues about how to unify all the forces in
nature, and has also tightened up our understanding of how field and
string theories are related, where the latter can be used to link
various field theories together.

We conclude by stressing that the double copy is a relatively recent
discovery, and that many aspects of this fascinating correspondence
remain to be explored. Given the wide range of physical phenomena that
are linked by the double copy, it is certainly possible for
researchers from a wide range of fields to actively contribute to
ongoing efforts in this area!

\section*{Author biography}

\begin{minipage}{0.2\textwidth}
\scalebox{0.075}{\includegraphics[angle=-90,origin=c]{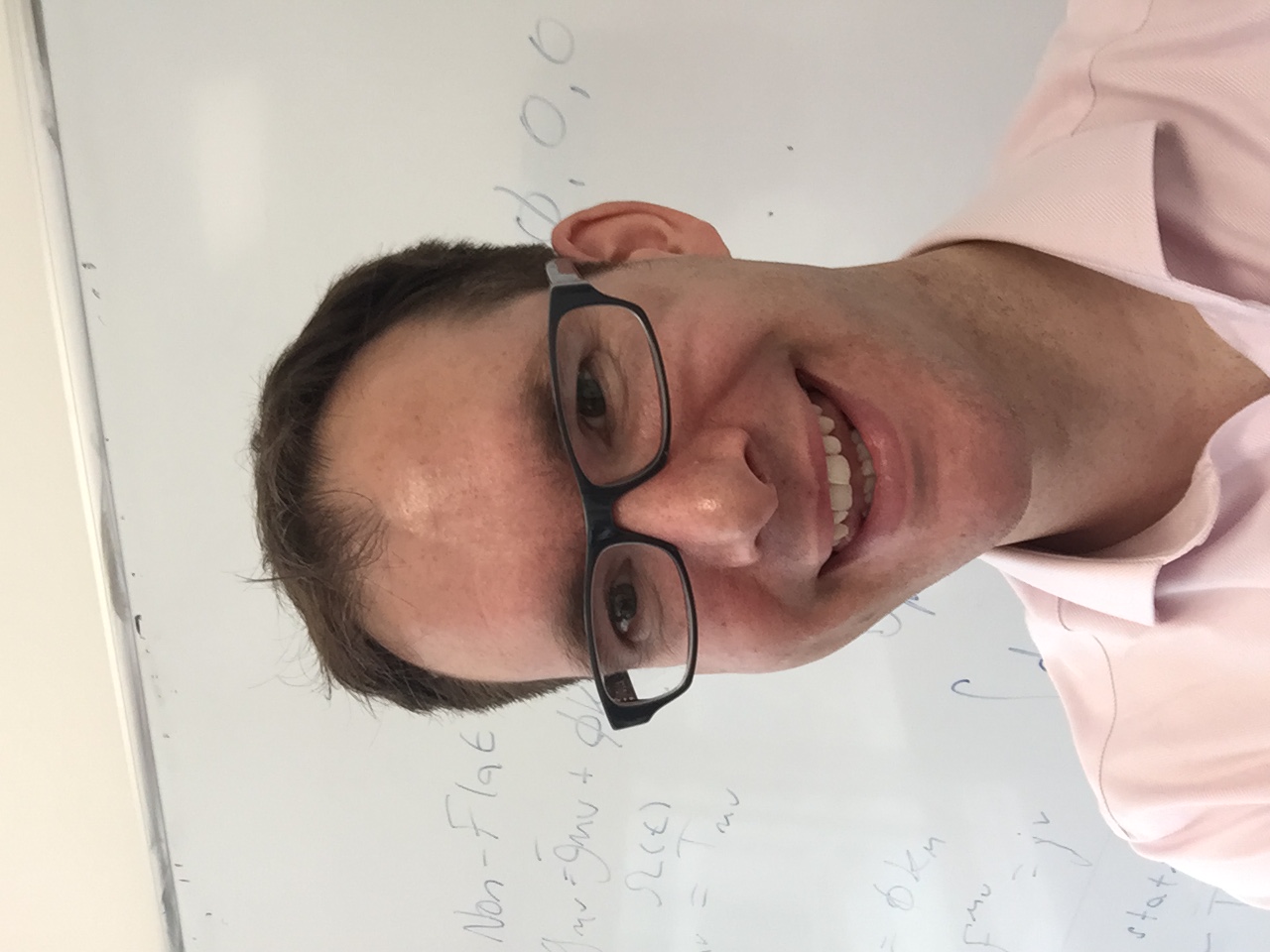}}
\end{minipage}
\hfill
\begin{minipage}{0.8\textwidth}
\begin{small}
Dr. Chris White is a Reader in Theoretical Physics at the Centre for
Research in String Theory, Queen Mary University of London. He
completed his PhD on the structure of the proton at high energy at the
University of Cambridge in 2006, followed by research appointments at
the Dutch National Institute for Nuclear and High Energy Physics
(Nikhef) in Amsterdam, and Durham University. He then held a
lectureship at the University of Glasgow, before moving to London in
2016. Chris is currently working on applications of Quantum
Chromodynamics (the theory of quarks and gluons) to collider
experiments, as well as how to relate this theory to gravity.
\end{small}
\end{minipage}

\section*{Acknowledgments}

CDW is very grateful to Nadia Bahjat-Abbas, Andr\'{e}s Luna, Stacey
Melville, Ricardo Monteiro, Isobel Nicholson, Donal O'Connell, Alex
Ochirov and Nicolas Westerberg for discussions and collaboration on
related topics. He is supported by the UK Science and Technology
Facilities Council (STFC).

\bibliographystyle{tfnlm} \bibliography{refs}

\begin{thebibliography}{10}
\providecommand{\url}[1]{\normalfont{#1}}
\providecommand{\urlprefix}{Available from: }

\bibitem{Abbott:2016blz}
Abbott~BP, et~al. (Virgo, LIGO Scientific). {Observation of Gravitational Waves
  from a Binary Black Hole Merger}. Phys Rev Lett.
  2016;\hspace{0pt}116(6):061102.

\bibitem{Weinberg:1980gg}
Weinberg~S. {U}ltraviolet divergences in quantum theories of gravitation. In:
  General relativity: An einstein centenary survey; 1980. p. 790--831.

\bibitem{Bern:2010ue}
Bern~Z, Carrasco~JJM, Johansson~H. {Perturbative Quantum Gravity as a Double
  Copy of Gauge Theory}. PhysRevLett. 2010;\hspace{0pt}105:061602.

\bibitem{Bern:2010yg}
Bern~Z, Dennen~T, Huang~Yt, et~al. {Gravity as the Square of Gauge Theory}.
  PhysRev. 2010;\hspace{0pt}D82:065003.

\bibitem{Bern:2008qj}
Bern~Z, Carrasco~JJM, Johansson~H. {New Relations for Gauge-Theory Amplitudes}.
  Phys Rev. 2008;\hspace{0pt}D78:085011.

\bibitem{Dirac:1928hu}
Dirac~PAM. {The quantum theory of the electron}. Proc Roy Soc Lond.
  1928;\hspace{0pt}A117:610--624.

\bibitem{Noether:1918zz}
Noether~E. {Invariant Variation Problems}. Gott Nachr.
  1918;\hspace{0pt}1918:235--257. [Transp. Theory Statist. Phys.1,186(1971)].

\bibitem{Feynman:1963ax}
Feynman~RP. {Quantum theory of gravitation}. Acta Phys Polon.
  1963;\hspace{0pt}24:697--722.

\bibitem{DeWitt:1967yk}
DeWitt~BS. {Quantum Theory of Gravity. 1. The Canonical Theory}. Phys Rev.
  1967;\hspace{0pt}160:1113--1148.

\bibitem{DeWitt:1967ub}
DeWitt~BS. {Quantum Theory of Gravity. 2. The Manifestly Covariant Theory}.
  Phys Rev. 1967;\hspace{0pt}162:1195--1239.

\bibitem{DeWitt:1967uc}
DeWitt~BS. {Quantum Theory of Gravity. 3. Applications of the Covariant
  Theory}. Phys Rev. 1967;\hspace{0pt}162:1239--1256.

\bibitem{Elvang:2013cua}
Elvang~H, Huang~Yt. {Scattering Amplitudes}. 2013;\hspace{0pt}.

\bibitem{Cheung:2017pzi}
Cheung~C. {TASI Lectures on Scattering Amplitudes}. 2017;\hspace{0pt}.

\bibitem{Bern:2012uf}
Bern~Z, Carrasco~JJM, Dixon~LJ, et~al. {Simplifying Multiloop Integrands and
  Ultraviolet Divergences of Gauge Theory and Gravity Amplitudes}. Phys Rev.
  2012;\hspace{0pt}D85:105014.

\bibitem{Bern:2013yya}
Bern~Z, Davies~S, Dennen~T, et~al. {Color-Kinematics Duality for Pure
  Yang-Mills and Gravity at One and Two Loops}. Phys Rev.
  2015;\hspace{0pt}D92(4):045041.

\bibitem{Monteiro:2011pc}
Monteiro~R, O'Connell~D. {The Kinematic Algebra From the Self-Dual Sector}.
  JHEP. 2011;\hspace{0pt}07:007.

\bibitem{Badger:2015lda}
Badger~S, Mogull~G, Ochirov~A, et~al. {A Complete Two-Loop, Five-Gluon Helicity
  Amplitude in Yang-Mills Theory}. JHEP. 2015;\hspace{0pt}10:064.

\bibitem{Saotome:2012vy}
Saotome~R, Akhoury~R. {Relationship Between Gravity and Gauge Scattering in the
  High Energy Limit}. JHEP. 2013;\hspace{0pt}1301:123.

\bibitem{Vera:2012ds}
Sabio~Vera~A, Serna~Campillo~E, Vazquez-Mozo~MA. {Color-Kinematics Duality and
  the Regge Limit of Inelastic Amplitudes}. JHEP. 2013;\hspace{0pt}1304:086.

\bibitem{Johansson:2013nsa}
Johansson~H, Sabio~Vera~A, Serna~Campillo~E, et~al. {Color-Kinematics Duality
  in Multi-Regge Kinematics and Dimensional Reduction}. JHEP.
  2013;\hspace{0pt}1310:215.

\bibitem{Johansson:2013aca}
Johansson~H, Sabio~Vera~A, Serna~Campillo~E, et~al. {Color-kinematics duality
  and dimensional reduction for graviton emission in Regge limit}.
  2013;\hspace{0pt}.

\bibitem{Melville:2013qca}
Melville~S, Naculich~SG, Schnitzer~HJ, et~al. {Wilson line approach to gravity
  in the high energy limit}. Phys Rev. 2014;\hspace{0pt}D89(2):025009.

\bibitem{Luna:2016idw}
Luna~A, Melville~S, Naculich~SG, et~al. {Next-to-soft corrections to high
  energy scattering in QCD and gravity}. JHEP. 2017;\hspace{0pt}01:052.

\bibitem{Weinberg:1965nx}
Weinberg~S. {Infrared photons and gravitons}. Phys Rev.
  1965;\hspace{0pt}140:B516--B524.

\bibitem{Naculich:2011ry}
Naculich~SG, Schnitzer~HJ. {Eikonal methods applied to gravitational scattering
  amplitudes}. JHEP. 2011;\hspace{0pt}05:087.

\bibitem{Laenen:2008gt}
Laenen~E, Stavenga~G, White~CD. {Path integral approach to eikonal and
  next-to-eikonal exponentiation}. JHEP. 2009;\hspace{0pt}03:054.

\bibitem{White:2011yy}
White~CD. {Factorization Properties of Soft Graviton Amplitudes}. JHEP.
  2011;\hspace{0pt}05:060.

\bibitem{Akhoury:2011kq}
Akhoury~R, Saotome~R, Sterman~G. {Collinear and Soft Divergences in
  Perturbative Quantum Gravity}. Phys Rev. 2011;\hspace{0pt}D84:104040.

\bibitem{Beneke:2012xa}
Beneke~M, Kirilin~G. {Soft-collinear gravity}. JHEP. 2012;\hspace{0pt}09:066.

\bibitem{Oxburgh:2012zr}
Oxburgh~S, White~C. {BCJ duality and the double copy in the soft limit}. JHEP.
  2013;\hspace{0pt}1302:127.

\bibitem{Johansson:2017srf}
Johansson~H, Nohle~J. {Conformal Gravity from Gauge Theory}. 2017;\hspace{0pt}.

\bibitem{Anastasiou:2017nsz}
Anastasiou~A, Borsten~L, Duff~MJ, et~al. {Are all supergravity theories
  Yang-Mills squared?}. 2017;\hspace{0pt}.

\bibitem{Borsten:2017jpt}
Borsten~L. {On $D=6$, $\mathcal{N}=(2,0)$ and $\mathcal{N}=(4,0)$ theories}.
  2017;\hspace{0pt}.

\bibitem{Cheung:2017ems}
Cheung~C, Shen~CH, Wen~C. {Unifying Relations for Scattering Amplitudes}.
  2017;\hspace{0pt}.

\bibitem{Kawai:1985xq}
Kawai~H, Lewellen~D, Tye~S. {A Relation Between Tree Amplitudes of Closed and
  Open Strings}. NuclPhys. 1986;\hspace{0pt}B269:1.

\bibitem{Cachazo:2013iea}
Cachazo~F, He~S, Yuan~EY. {Scattering of Massless Particles: Scalars, Gluons
  and Gravitons}. 2013;\hspace{0pt}.

\bibitem{Cachazo:2013hca}
Cachazo~F, He~S, Yuan~EY. {Scattering of Massless Particles in Arbitrary
  Dimension}. 2013;\hspace{0pt}.

\bibitem{Cachazo:2013gna}
Cachazo~F, He~S, Yuan~EY. {Scattering Equations and KLT Orthogonality}.
  2013;\hspace{0pt}.

\bibitem{Mason:2013sva}
Mason~L, Skinner~D. {Ambitwistor strings and the scattering equations}. JHEP.
  2014;\hspace{0pt}1407:048.

\bibitem{Geyer:2014fka}
Geyer~Y, Lipstein~AE, Mason~LJ. {Ambitwistor strings in 4-dimensions}.
  PhysRevLett. 2014;\hspace{0pt}113:081602.

\bibitem{Casali:2015vta}
Casali~E, Geyer~Y, Mason~L, et~al. {New Ambitwistor String Theories}. JHEP.
  2015;\hspace{0pt}11:038.

\bibitem{Casali:2016atr}
Casali~E, Tourkine~P. {On the null origin of the ambitwistor string}. JHEP.
  2016;\hspace{0pt}11:036.

\bibitem{Geyer:2015bja}
Geyer~Y, Mason~L, Monteiro~R, et~al. {Loop Integrands for Scattering Amplitudes
  from the Riemann Sphere}. Phys Rev Lett. 2015;\hspace{0pt}115(12):121603.

\bibitem{Geyer:2015jch}
Geyer~Y, Mason~L, Monteiro~R, et~al. {One-loop amplitudes on the Riemann
  sphere}. JHEP. 2016;\hspace{0pt}03:114.

\bibitem{Cheung:2016say}
Cheung~C, Remmen~GN. {Twofold Symmetries of the Pure Gravity Action}. JHEP.
  2017;\hspace{0pt}01:104.

\bibitem{Bern:1999ji}
Bern~Z, Grant~AK. {Perturbative gravity from QCD amplitudes}. Phys Lett.
  1999;\hspace{0pt}B457:23--32.

\bibitem{Monteiro:2014cda}
Monteiro~R, O'Connell~D, White~CD. {Black holes and the double copy}. JHEP.
  2014;\hspace{0pt}1412:056.

\bibitem{Tangherlini:1963bw}
Tangherlini~F. {Schwarzschild field in n dimensions and the dimensionality of
  space problem}. Nuovo Cim. 1963;\hspace{0pt}27:636--651.

\bibitem{Myers:1986un}
Myers~RC, Perry~M. {Black Holes in Higher Dimensional Space-Times}. Annals
  Phys. 1986;\hspace{0pt}172:304.

\bibitem{Ridgway:2015fdl}
Ridgway~AK, Wise~MB. {Static Spherically Symmetric Kerr-Schild Metrics and
  Implications for the Classical Double Copy}. 2015;\hspace{0pt}.

\bibitem{Chong:2004hw}
Chong~Z, Gibbons~G, Lu~H, et~al. {Separability and killing tensors in
  Kerr-Taub-NUT-de sitter metrics in higher dimensions}. PhysLett.
  2005;\hspace{0pt}B609:124--132.

\bibitem{Taub}
Taub~AH. Empty space-times admitting a three parameter group of motions. Annals
  of Mathematics. 1951;\hspace{0pt}53(3):pp. 472--490.
  \urlprefix\url{http://www.jstor.org/stable/1969567}.

\bibitem{NUT}
Newman~E, Tamburino~L, Unti~T. Empty‐space generalization of the
  schwarzschild metric. Journal of Mathematical Physics.
  1963;\hspace{0pt}4(7):915--923.
  \urlprefix\url{http://scitation.aip.org/content/aip/journal/jmp/4/7/10.1063/1.1704018}.

\bibitem{Luna:2015paa}
Luna~A, Monteiro~R, O'Connell~D, et~al. {The classical double copy for
  Taub–NUT spacetime}. Phys Lett. 2015;\hspace{0pt}B750:272--277.

\bibitem{Coleman:1977ps}
Coleman~SR. {Nonabelian Plane Waves}. PhysLett. 1977;\hspace{0pt}B70:59.

\bibitem{Aichelburg:1970dh}
Aichelburg~P, Sexl~R. {On the Gravitational field of a massless particle}.
  GenRelGrav. 1971;\hspace{0pt}2:303--312.

\bibitem{Siegel:1999ew}
Siegel~W. {Fields}. 1999;\hspace{0pt}.

\bibitem{Barnett:2014era}
Barnett~SM. {Maxwellian theory of gravitational waves and their mechanical
  properties}. New JPhys. 2014;\hspace{0pt}16:023027.

\bibitem{Adamo:2017nia}
Adamo~T, Casali~E, Mason~L, et~al. {Scattering on plane waves and the double
  copy}. 2017;\hspace{0pt}.

\bibitem{Luna:2016due}
Luna~A, Monteiro~R, Nicholson~I, et~al. {The double copy: Bremsstrahlung and
  accelerating black holes}. 2016;\hspace{0pt}.

\bibitem{Stephani:2003tm}
Stephani~H, Kramer~D, MacCallum~MA, et~al. {Exact solutions of Einstein's field
  equations}. 2003;\hspace{0pt}.

\bibitem{Goldberger:2016iau}
Goldberger~WD, Ridgway~AK. {Radiation and the classical double copy for color
  charges}. Phys Rev. 2017;\hspace{0pt}D95(12):125010.

\bibitem{Goldberger:2017frp}
Goldberger~WD, Prabhu~SG, Thompson~JO. {Classical gluon and graviton radiation
  from the bi-adjoint scalar double copy}. 2017;\hspace{0pt}.

\bibitem{Luna:2016hge}
Luna~A, Monteiro~R, Nicholson~I, et~al. {Perturbative spacetimes from
  Yang-Mills theory}. JHEP. 2017;\hspace{0pt}04:069.

\bibitem{Bjerrum-Bohr:2013bxa}
Bjerrum-Bohr~NEJ, Donoghue~JF, Vanhove~P. {On-shell Techniques and Universal
  Results in Quantum Gravity}. JHEP. 2014;\hspace{0pt}02:111.

\bibitem{Bjerrum-Bohr:2016hpa}
Bjerrum-Bohr~NEJ, Donoghue~JF, Holstein~BR, et~al. {Light-like Scattering in
  Quantum Gravity}. JHEP. 2016;\hspace{0pt}11:117.

\bibitem{Anastasiou:2014qba}
Anastasiou~A, Borsten~L, Duff~MJ, et~al. {Yang-Mills origin of gravitational
  symmetries}. Phys Rev Lett. 2014;\hspace{0pt}113(23):231606.

\bibitem{Anastasiou:2016csv}
Anastasiou~A, Borsten~L, Duff~MJ, et~al. {Twin supergravities from Yang-Mills
  theory squared}. Phys Rev. 2017;\hspace{0pt}D96(2):026013.

\bibitem{White:2016jzc}
White~CD. {Exact solutions for the biadjoint scalar field}. Phys Lett.
  2016;\hspace{0pt}B763:365--369.

\bibitem{DeSmet:2017rve}
De~Smet~PJ, White~CD. {Extended solutions for the biadjoint scalar field}.
  2017;\hspace{0pt}.

\bibitem{Wu:1967vp}
Wu~TT, Yang~CN. Some solutions of the classical isotopic gauge field equations.
  1967;\hspace{0pt}.

\bibitem{Bern:2007hh}
Bern~Z, Carrasco~JJ, Dixon~LJ, et~al. {Three-Loop Superfiniteness of N=8
  Supergravity}. Phys Rev Lett. 2007;\hspace{0pt}98:161303.

\bibitem{Bern:2009kd}
Bern~Z, Carrasco~JJ, Dixon~LJ, et~al. {The Ultraviolet Behavior of N=8
  Supergravity at Four Loops}. Phys Rev Lett. 2009;\hspace{0pt}103:081301.

\end{thebibliography}

\end{document}